\documentclass[
  english,        
  font=times,     
  onecolumn,      
]{tumarticle}

\usepackage{lipsum}
\usepackage{dirtytalk}
\usepackage{amsmath}
\usepackage{hyperref}
\usepackage{dblfloatfix}
\usepackage{color}
\usepackage{placeins}
\usepackage{lineno}
\usepackage[backend=biber,style=authoryear]{biblatex}
\title{2D reaction-diffusion model-based biopsy simulation for dynamic tumor growth parameter estimation}

\author[affil=1, email=veronika.anja.hofmann@tum.de, orcid=0009-0003-5237-7133]{Veronika Hofmann}
\author[affil=2 3 4, orcid=0000-0001-9619-728X]{Pirmin Schlicke}
\author[affil=1, orcid=0000-0003-1602-2936]{Jan Zawallich}
\author[affil=2 3, orcid=0000-0002-9696-6410]{Heiko Enderling}
\author[affil=1, orcid=0000-0002-5521-4159]{Christina Kuttler}

\affil[mark=1]{\theDepartmentName, \theUniversityName, Munich, Germany}
\affil[mark=2]{Department of Radiation Oncology, The University of Texas MD Anderson Cancer Center, Houston, Texas, USA}
\affil[mark=3]{Institute for Data Science in Oncology, The University of Texas MD Anderson Cancer Center, Houston, Texas, USA}
\affil[mark=4]{Department of Biosciences and Medical Biology, University of Salzburg, Salzburg, Austria}

\date{September 2026}

\begin{document}

\maketitle

\begin{abstract}
  Once diagnosed, cancer requires a fast, reliable and preferably cost-efficient assessment of the current state and potential progression of the disease. A new method for estimating tumor cell diffusivity $D$ and proliferation rate $\gamma$ in the context of the mechanistic reaction-diffusion equation from single-point-in-time routine biopsies aims to deliver just that, and quantities computed from the parameter estimates have recently been tested as a new biomarkers for risk-stratification in radiotherapy. Here, we extend the findings of this previous work by providing a first theoretical validation. The method is applied to \textit{in silico} biopsies which are generated by solving the two-dimensional reaction-diffusion equation for different growth terms (exponential and logistic) with a Dirac-Delta initial condition, and transforming the continuous results into spatial point patterns via a form of reverse coarse-graining. If no information about tumor age is used, in short-term experiments the original dispersion length $\sqrt{D/\gamma}$ could be retrieved with a relative root mean squared error (RRMSE) of around $8\%$ and an $\text{R}^2$-value of 0.97.
  In long-term experiments, the RRMSEs ranged from 8 to $14\%$
  and the $\text{R}^2$-values from 0.75 to 0.98.
  The scaled front velocity $\sqrt{D \cdot\gamma}$, which can only be estimated if information about tumor age is available, was retrieved with an RRMSE of $7\%$ and an $\text{R}^2$ of 0.98 in both, the short-term and the long-term experiments.
\end{abstract}


\section*{Statements and Declarations}
\paragraph{Competing interests.}
The authors have no competing interests to declare that are relevant to the content of this article.

\section{Introduction}
The spectral-spatial analysis (SSA) method for the estimation of solid tumor growth and diffusion parameters based on a single-point-in-time haematoxylin and eosin (H\&E) stained biopsy has recently been developed~(\cite{Pasetto.2024, PirminsNewPaper}). The method computes the two-point correlation function (2PCF) from the nucleus-induced point pattern in the biopsy, transforms it into its power spectral density (PSD), and performs a curve fit with the two functions and their counterparts from a reaction-diffusion (RD) model for tumor growth.
The resulting parameter estimates for the diffusion coefficient $D$ and the unit growth $\gamma$ are combined into two biomarkers for individual survival prognosis: the dispersion length $\sqrt{D/\gamma}$ and the front velocity $2\sqrt{D\cdot\gamma}$, which have proven successful in risk stratification~(\cite{PirminsNewPaper}).
While prognostic tools such as the SSA of H\&E stained biopsies offer great potential given that the necessary data is routinely available at cancer diagnosis~(\cite{cancerdiagnosiswebsite}), new biomarkers tend to not be easily adapted by the medical practitioners, especially when they originate from non-medical fields such as mathematical modeling~(\cite{Yankeelov.2013, Anderson.2008, Eldabi.2009}). Two main reasons for this are that models tend to be calibrated and validated not thoroughly enough and hence do not appear trustworthy,
and that their calibration requires data that is rarely obtained in clinical routine care~(\cite{Yankeelov.2013}). As mentioned, the latter is not really an issue with the method at hand, given that H\&E stained biopsies are standard protocol for most cancer diagnoses, but the preoccupations arising from the former need to be addressed.

To achieve that, this work is the second pillar of a two-fold validation: the first was the aforementioned clinical investigation of the biomarkers, and here we are concerned with a mathematical \say{sanity check} which examines how well the SSA method can estimate these quantitative biomarkers under laboratory conditions.

On the way towards this validation we need to construct a biopsy simulation which is based on a continuum model for tumor growth, the RD-equation. Such a simulated \textit{in silico} biopsy offers the kind of controlled environment that is necessary to compare input and output growth and diffusion parameters. Hence, this work offers more than the validation protocol of the SSA method: we also demonstrate how the two-dimensional RD-equation can be solved for exponential and logistic growth terms with a numerically approximated Dirac-Delta initial condition (i.c.), and we present an algorithm that allows to create \textit{in silico} biopsies from continuous cell densities, together with a method for normalization of such densities.

\subsection*{Related Work}
While there are several studies on how the (multi-dimensional) Dirac-Delta function can be numerically approximated if it serves as a source term \textit{in} a partial differential equation (PDE) (for instance~\cite{Tornberg.2004, Engquist.2005, DiSchianoCola.2021, Ashyraliyev.2008}), investigations regarding a discretized Dirac-Delta as an i.c.~are rare, especially in a polar/spherical geometry which will be used here. 
However, the source-point nature of the Dirac-Delta allows for easier analytical computations in the phase space than other usual i.c.s, such as a Gaussian curve. 
The PSD that is used for the SSA curve fit is set up from an RD-equation with exponential growth and a Dirac-Delta initial term, hence to be able to assess its growth type-dependent performance, it is adequate to solve the model PDEs on which the \textit{in silico} biopsies are based with the same initial function. For later evaluations it is of course possible to challenge the method further and use alternative i.c.s, but for a first assessment we should keep close to the theoretical conditions to avoid a false-negative result.

The typical method of choice for the generation of \textit{in silico} biopsies would be agent-based models (ABM) or cellular automata. These kinds of models use the individual needs and interests of every single cell (or agent) as \say{force} that dictates the system's dynamics. While ABMs and cellular automata can describe reaction-diffusion dynamics well and are widely used in mathematical oncology (see~\cite{Stephan.2024} for a general review, and \cite{Hutchinson.2022, Somer.2026} for examples of ABM biopsy simulations), they tend to be relatively slow: in every time step, the actions of a growing number of cells need to be determined, leading to a lower bound of algorithm speed of $\mathcal{O}(t \cdot n)$, where $t$ is the number of time steps and $n$ is the average number of cells in the system at a given time step. If interactions between the cells need to be considered as well, we quickly arrive at an algorithm complexity of $\mathcal{O}(n^2 \cdot t)$ or worse. Using a continuum model to first compute cell densities at the desired final point in time, and only then place the cells on the domain will cut the computational expense to $\mathcal{O}(t \cdot m)$ for solving the model equation with $t$ time steps and $m$ spatial nodes plus $\mathcal{O}(n)$ or $\mathcal{O}(n^2)$ for the cell placement, i.e.,~order of $\mathcal{O}(n^2 + t\cdot m)$ time complexity. The number $m$ can be kept small by taking advantage of the geometry of the domain and the properties of the model (e.g., a radial symmetry, or a decreasing concentration allowing to use a lower resolution further away from the tumor's core). Also, with this two-sided approach it only takes order $\mathcal{O}(r \cdot n^2)$ computational expense to create $r$ realizations of a simulated biopsy because the solution of the continuum model only needs to be computed once -- in comparison to $\mathcal{O}(r \cdot n^2 \cdot t)$ needed by purely stochastic models. Apart from the runtime argument, it should be mentioned that using the continuous RD-model bears the advantage that we can profit from the existing research on its properties.
The methodology proposed in this work can be understood as a kind of \say{reverse coarse graining}. While coarse graining (i.e.,~convert an ABM to a continuous model) is relatively common in mathematical biology (see~\cite{Saunders.2013} for a review, and~\cite{Martinson.2024} for a recent quantitative analysis), the transition from a continuous model to a discrete distribution is a less thoroughly explored topic, especially in the case of biological application. Similar procedures can be found in the field of molecular dynamics, where the method is mainly used to reduce computational costs of simulations~(\cite{Rosenberger.2016, AbiMansour.2016, Wassenaar.2014}).

\begin{figure}[htb] 
\centering 
\includegraphics[width=\columnwidth]{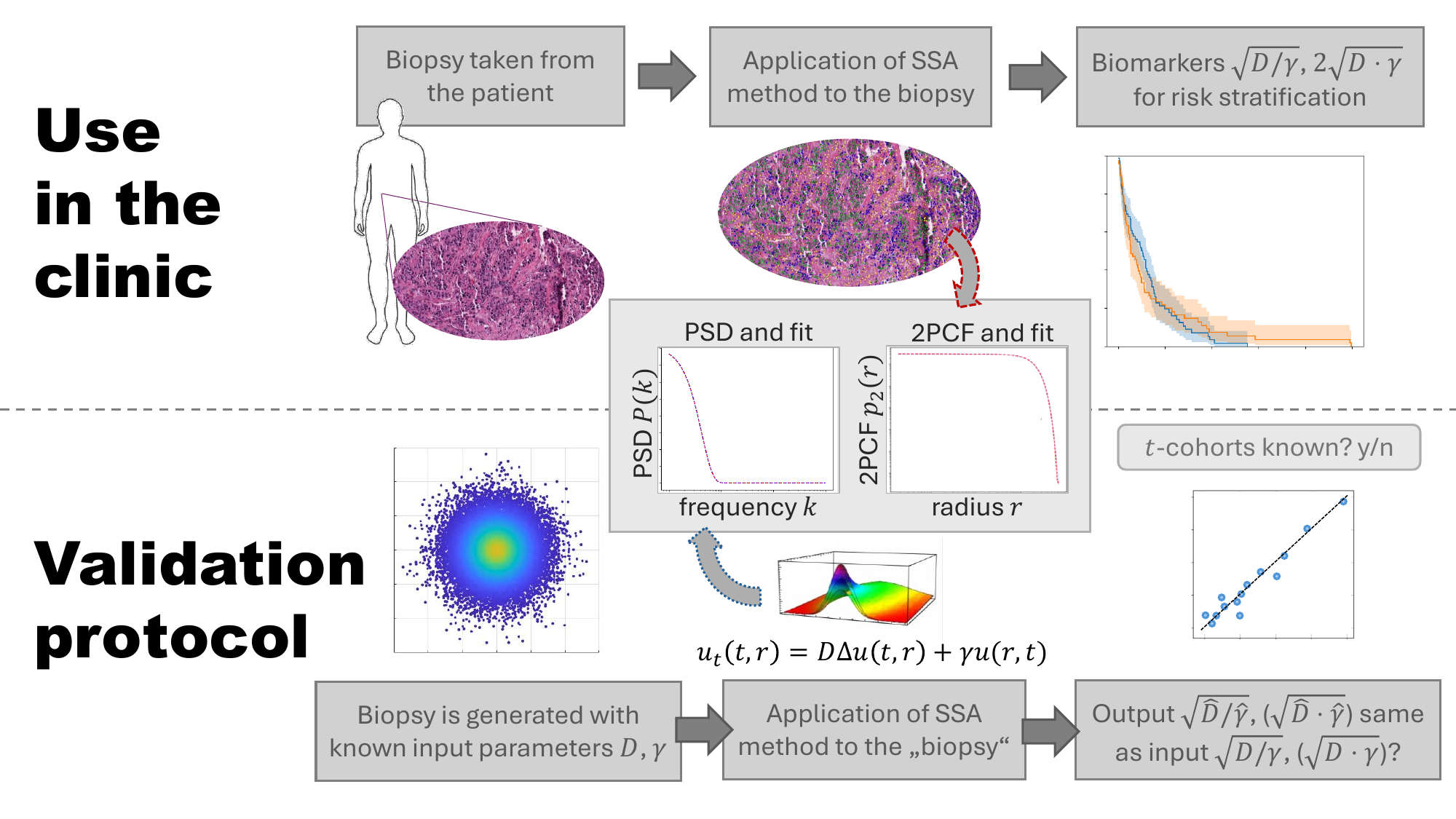} 
\caption{\label{fig:validation_procedure}Schematic depiction of the targeted clinical use of the SSA method (top row) together with the \textit{in silico} validation process that is performed in this work (bottom row). Note that while \cite{PirminsNewPaper} use the biomarker $2\sqrt{D\cdot\gamma}$, here we consider the expression $\sqrt{D\cdot \gamma}$ without the factor.}
\end{figure}

In the literature, the term \say{\textit{in silico} validation} of biomarkers has different meanings. 
It can describe the process of biomarker discovery by applying statistical methods to large data sets and extracting significant accumulations of certain proteins or genes, which are then the disease biomarkers (e.g.~\cite{Gomez.2023}). In a sense, this is discovery and validation in a single step, where the quality of the validation highly depends on the size and the quality of the data set. 
The term can also be used for the scenario where the biomarker consists of the results of a mathematical model, which is validated with patient data by using the model to create digital twins of these patients (e.g.~\cite{Swanson.2008, Hutchinson.2022}). Properly calibrated digital twins can be model realizations that use the individual features of a patient as input data, and potentially allow to make predictions on the individual disease progression. If the model predictions agree with the observed outcomes, the validation is successful.
In this paper, \textit{in silico} validation means that the biomarker results from a mathematical model-based method, the SSA, and we test it by applying the method to simulated data with known underlying parameters that are to be recovered (e.g.~\cite{Somer.2026}; see Fig.~\ref{fig:validation_procedure} for an illustration of the process). This is necessarily different from the digital twins approach because it aims to validate the method on untreated tissue, a circumstance that one hardly encounters in data from human patients.

\section{Background}

In this section we introduce the notation and general properties that we will use throughout this paper.

The continuum equation on which both the SSA method and the simulated biopsies are based is an RD-equation with exponential growth term. In two spatial dimensions, the evolution of the cell density $u$ is assumed to be radially symmetric around the origin at a given point in time $t \in [0, t_\text{max}]$, and can be written in polar coordinates $(r, \phi) \in [0, r_\text{max}] \times [0, 2\pi)$ as
\begin{equation}\label{eq:2D_polar_RD_exp_growth}
    \partial_t u(r, t) = D\Big(\partial_r^2 u (r, t) + \frac{1}{r} \partial_r u(r, t)\Big) + \gamma u(r, t),
\end{equation}
where the angular component $\phi$ is already omitted due to symmetry. $D > 0$ is the diffusion coefficient and its unit can be chosen, e.g., as $\frac{\mu\text{m}^2}{\text{day}}$, and $\gamma > 0$ represents the growth rate, for instance given in $\frac{1}{\text{day}}$. It is also possible to choose more general units such as $\frac{\text{area}}{\text{time}}$ and $\frac{1}{\text{time}}$. In this work, the precise choice of units is not critical and has a rather illustrative purpose. Alternatively to the exponential growth, \eqref{eq:2D_polar_RD_exp_growth} can also feature a logistic growth term leading to
\begin{align}\label{eq:2D_polar_RD_log_growth}
\begin{split}
    \partial_t u(r, t) = D\Big(\partial_r^2 u (r, t) + \frac{1}{r} \partial_r u(r, t)\Big) + \gamma u(r, t) \left( 1 - u(r, t) \right),
\end{split}
\end{align}
which is valid in this form for dimensionless $u$. To make it clear which of the RD-equations \eqref{eq:2D_polar_RD_exp_growth} and \eqref{eq:2D_polar_RD_log_growth} we discuss, we will refer to their solutions as $u_\text{exp}$ and $u_\text{log}$, and if the statement concerns both we will simply write $u$.
For both equations, we use a two-dimensional Dirac-Delta distribution as i.c.: if $\delta: \mathbb{R} \to \mathbb{R}$ is a function approximating the Dirac-Delta distribution, then we use
\begin{equation*}
    u(r, 0) = \frac{1}{2 \pi r} \delta(r) =: \delta^2(r),
\end{equation*}
$\delta^2: \mathbb{R}\to \mathbb{R}$, as an approximation of the two-dimensional Dirac-Delta. See appendix~\ref{app:dirac_delta_derivation} for a derivation. We have two types of boundary conditions (b.c.s): Neumann b.c.s at $r=0$ to ensure the radial symmetry, i.e.,
\begin{equation*}
    \partial_r u(0, t) = 0,
\end{equation*}
and Dirichlet b.c.s at $r = r_{\text{max}}$ to obtain a unique, well-defined solution:
\begin{equation*}
    u(r_{\text{max}}, t) = 0.
\end{equation*}
For \eqref{eq:2D_polar_RD_exp_growth} with the above i.c.~and b.c.s, an analytical solution exists if we let $r_{\text{max}} \to \infty$ (see for instance~\cite{Josza.2013}): 
\begin{equation}\label{eq:analytical_sol_2D_exp}
    u_\text{exp} (r, t) = \frac{1}{4 \pi D t} e^{-\frac{r^2}{4Dt} + \gamma t}.
\end{equation}

The RD-equations are discretized equidistantly in time and space on one-dimensional grids: $u_i^n$ is the cell density at radius $(r_i)_{i \in \{1, ..., I\}} = i \cdot \Delta r$, and time $(t^n)_{n \in \{1, ..., N\}} = n \cdot \Delta t$, where $\Delta r$ and $\Delta t$ are the distances between the space- and time-nodes, respectively. This uniform version of the spatial grid will be defined as $R_\text{unif}$, in contrast to several other non-uniform variants that are tested in this work to see whether local grid refinements close to the origin improve performance. The Dirac-Delta i.c.~theoretically has an infinite rate of change at $r=0$, bearing a source of stiffness to the PDE which we will try to mitigate by using the following grid stretchings:

\begin{itemize}
    \item $R_\text{unif}$: $\{ 0, \Delta r, 2\Delta r, ..., r_\text{max} \}$ (for reference)
    \item $R_1$: use $R_\text{unif}$ for $r \in [1, r_\text{max}]$, and uniformly increase the number of points between 0 and 1 by a factor 10, i.e., $\{ 0, \frac{\Delta r}{10}, 2\frac{\Delta r}{10}, ..., 9\frac{\Delta r}{10}, \Delta r, \Delta r+\frac{\Delta r}{10}, ..., 1, 1+ \Delta r, 1+2\Delta r, ..., r_\text{max} \}$
    \item $R_2$: use $R_\text{unif}$ for $r \in [1, r_\text{max}]$, and transform $R_\text{unif}$ between 0 and 1 by using $R_\text{unif}^3$, i.e., \\
    $\{ 0, (\Delta r)^3, (2\Delta r)^3, ..., 1, 1 + \Delta r, 1+2\Delta r, ...,  r_\text{max} \}$
    \item $R_3$: use $\frac{1}{r_\text{max}} R_\text{unif}^2$ in the whole domain., i.e., $\frac{1}{r_\text{max}}\cdot\{ 0, (\Delta r)^2, (2\Delta r)^2, ..., r_\text{max} \}$
    \item $R_4$: use $\frac{1}{r_\text{max}^2} R_\text{unif}^3$ in the whole domain, i.e., $\frac{1}{r_\text{max}^2}\cdot\{ 0, (\Delta r)^3, (2\Delta r)^3, ..., r_\text{max} \}$
\end{itemize}

For a given discretization $R_i$, $i = 1, ..., 4$, we define the vector of grid distances $(\Delta r_1, ..., \Delta r_{I-1})^T$ with $\Delta r_i = r_{i+1} - r_i$, then $r_i = \sum_{j=1}^{i-1} \Delta r_j$.

The quality of the numerical solution of~\eqref{eq:2D_polar_RD_log_growth} is evaluated with two measures: convergence speed and mass conservation. Convergence speed will be assessed in this work using two methods, either by comparing the numerical solution $u_\text{log}^{h,\text{fine}}$ on very fine spatial and temporal grids to solutions $u_\text{log}^h$ that were computed on less fine grids, or by applying the numerical solver to~\eqref{eq:2D_polar_RD_exp_growth}, where we can use the analytical solution instead of a highly-resolved numerical solution. The former comparison will yield information on how fast the solver converges for the equation of interest, and the latter will provide an estimate of any systematic errors of solver and discretization. The differences will be evaluated with the $L^2$-error, which we define by
\begin{equation*}
    \| u_\text{fine} - u_\text{coarse} \|_{L^2} = \left( 2\pi \int_0^{r_\text{max}} r(u_\text{fine} - u_\text{coarse})^2 \:\text{d}r \right)^{\frac{1}{2}},
\end{equation*}
where $u_\text{fine} = u_\text{fine}(r, t_\text{max})$ refers either to the numerical solution on the fine grid or to the analytical solution, and $u_\text{coarse} = u_\text{coarse}(r, t_\text{max})$ is the numerical solution on the coarser grid. Technically, $u_\text{fine}$ and $u_\text{coarse}$ are vectors and need to be evaluated discretely for each $r_i$; here, they are handled as functions for notational simplicity.

The mass conservation property of the solver in the reaction-free state ($\gamma = 0$) is particularly important because it is a finite difference solver, i.e.,~it is not guaranteed that the solver will preserve the cell density other than e.g.~finite volume methods. To measure it, the total cell density is computed at $t=0$ and $t=t_\text{max}$ and the two values are compared. Total cell density is calculated using the volume integral, i.e.,
\begin{align*}
    \int_{\mathbb{R}^2} u(x, t^n) \: \text{d}x &=  2\pi \int_0^\infty r\: u(r, t^n) \: \text{d} r, 
\end{align*}
which can be discretized using the trapezoidal rule on non-uniform grids, 
\begin{equation*}
    \int_a^b f(x) \:\text{d}x \approx \frac{1}{2} \sum_i (x_{i+1}-x_i) (f(x_i) + f(x_{i+1})),
\end{equation*} 
where $(x_i)_{i\in\{ 1, ..., I \}}$ are the spatial nodes between $a$ and $b$. Of course, this rule is only applicable if $u(r)$ approaches $0$ fast enough for $r\to\infty$, i.e., we need $u(r_\text{max}) \approx 0$. This requires appropriate choices for the parameters $r_\text{max}$, $D$, $\gamma$: the domain needs to be large enough, and growth and diffusion need to be slow enough to ensure that the cells do not reach the boundaries of the domain within $t_\text{max}$.
For our discretized integral we conclude
\begin{align*}
    \int_{\mathbb{R}^2} u(x) \: \text{d}x \approx 2\pi \cdot \frac{1}{2} \sum_{i=1}^{I-1} \Delta r_i\:(u_i^nr_i + u_{i+1}^nr_{i+1}).
\end{align*}

The SSA method uses two functions in its parameter estimation algorithm that are fitted against the data from the biopsy, the two-point correlation function (2PCF) and the power spectral density (PSD). While we will not go into detail here and instead refer to~\cite{Pasetto.2024, PirminsNewPaper}, we want to mention that the goodness-of-fit measure, which serves as termination condition for the SSA fitting algorithm and will also be employed here for the assessment of the validation, is the coefficient of determination $\text{R}^2$, which we compute via
\begin{equation*}
    \text{R}^2 = 1 - \frac{\sum_{i=1}^n (y_i - f(\hat{y}_i))^2}{\sum_{i=1}^n(y_i - \bar{y})^2}.
\end{equation*}
Here, $y_1, ..., y_n$ are the observed values, $\hat{y}_1, ..., \hat{y}_n$ are the values from the fitting function, and $\bar{y}$ is the mean of the observations. We will stick to the $\hat{\cdot}$-notation for the validation as well, meaning that the true input parameter values are written as $D$, $\gamma$, and the estimated output values are $\hat{D}$, $\hat{\gamma}$.

Regarding goodness-of-fit, we will use an additional error measure for the validation, the relative root mean squared error (RRMSE) given by
\begin{equation*}
    \text{RRMSE} = \frac{\sqrt{\frac{1}{n}\sum_{i=1}^n (y_i - f(\hat{y}_i))^2}}{\bar{y}}.
\end{equation*}

This two-fold quality control yields the best of both worlds: with the RRMSE we make the quality of parameter estimations comparable between $D$, $\gamma$ and the biomarkers $\sqrt{D/\gamma}$, $\sqrt{D \cdot \gamma}$, as they cover different realistic ranges. With $\text{R}^2$, we aim to detect systemic errors in the estimation of a single parameter because the RRMSE can lead to an overestimation of the error in cases where the variation within the data is large. If the model still performs well in capturing the relationship between true parameter and estimate, $\text{R}^2$ will reveal it.

\section{Methods}

\subsection{PDE Solver}
Numerically solving equation \eqref{eq:2D_polar_RD_log_growth} consists of two steps: set up the solver, and find a suitable approximation for the Dirac-Delta i.c. As for the solver, we first revisit the spatial grid. All solutions need to progress from the initial Dirac-Delta function, which requires a high resolution close to $r = 0$ to prevent a loss of mass. Farther away from the origin, such a high resolution is unnecessary and will only increase computational cost, which is why several of our space discretizations augment precision only in the domain close to $r = 0$.

For the spatial discretization we choose a finite difference method for its simplicity in implementation, and a Crank-Nicolson scheme in time because of its second order convergence, matching the order of convergence in space. While more sophisticated methods such as finite volumes might at first glance offer a better alternative regarding robustness to the extreme i.c.~and therefore improve mass conservation and convergence, this impression is misleading: the error that stems from the Dirac-Delta approximation will be protruded in the form of the reaction term, leading to quadratic error propagation in the case of logistic growth. Hence even if the solver is performing well for a linear growth term, it is not guaranteed that it will be similarly reliable for a different type of growth.

Again, let $u_i^n$ denote the approximate solution of $u: [0, r_\text{max}] \times [0, t_\text{max}] \to \mathbb{R}$ at the $n$-th point in time and the $i$-th spatial point along the radius, then the scheme is given by (compare e.g.~to \cite{Jones.1980})

\begin{align*}
    \partial_t u(r, t) &\approx \frac{u_i^{n+1} - u_i^n}{\Delta t}, \\
    \partial_r u(r, t) &\approx \frac{1}{2} \left( \frac{u_{i+1}^{n+1} - u_{i-1}^{n+1}}{\Delta r_i + \Delta r_{i-1}} + \frac{u_{i+1}^{n} - u_{i-1}^{n}}{\Delta r_i + \Delta r_{i-1}} \right), \\
    \partial^2_r u(r, t) &\approx \frac{1}{\Delta r_i + \Delta r_{i-1}} \biggl( \frac{u_{i+1}^{n+1} - u_i^{n+1} + u_{i+1}^{n} - u_i^n}{\Delta r_i} - \frac{u_{i}^{n+1} - u_{i-1}^{n+1} + u_i^n - u_{i-1}^{n}}{\Delta r_{i-1}} \biggr),
\end{align*}
so that the governing equation of the solver results in
\begin{equation*}
    \begin{split}
        \frac{u_i^{n+1}}{\Delta t} - D \biggl( \frac{1}{\Delta r_i + \Delta r_{i-1}} \biggl( \frac{u_{i+1}^{n+1} - u_i^{n+1}}{\Delta r_i} - \frac{u_{i}^{n+1} - u_{i-1}^{n+1}}{\Delta r_{i-1}} \biggr) + \frac{1}{2r_i} \cdot \frac{u_{i+1}^{n+1} - u_{i-1}^{n+1}}{\Delta r_i + \Delta r_{i-1}} \biggr) - \frac{\gamma u_i^{n+1}}{2} \biggl( 1 - u_i^{n+1} \biggr) \\
        = \frac{u_i^{n}}{\Delta t} + D \biggl( \frac{1}{\Delta r_i + \Delta r_{i-1}} \biggl( \frac{u_{i+1}^{n} - u_i^{n}}{\Delta r_i} - \frac{u_{i}^{n} - u_{i-1}^{n}}{\Delta r_{i-1}} \biggr) + \frac{1}{2r_i} \cdot \frac{u_{i+1}^{n} - u_{i-1}^{n}}{\Delta r_i + \Delta r_{i-1}} \biggr) + \frac{\gamma u_i^n}{2} \biggl( 1 - u_i^n \biggr).
    \end{split}
\end{equation*}
The Dirichlet b.c.~is simply implemented by setting $u_I^n = 0$ for all times $n$.
The Neumann b.c.~is set by using the ghost point technique~(\cite{Morton.2012}) (see~\cite{Crossley.2023} for an exemplary application): we introduce a spatial node at $i=0$. 

As we started counting the nodes in the spatial domain at $i=1$, this point technically does not exist but we imagine it to lie just outside of the domain next to $u_1$, with $u_0 = u_2$ to describe the \say{mirroring} at $r_1$. This definition is possible because the Neumann b.c.~imposes a no-flux boundary, i.e.,~no cancer cells are entering or leaving the domain at $r=0$, a premise which we do not hurt if we extend the domain by this imaginary point. In the grid $R_\text{unif}$, the assumption $\Delta r_0 = \Delta r$ is indisputable, but in the non-uniform grids the only similarly straightforward definition $\Delta r_0 = \Delta r_1$ is not necessarily the most appropriate. For lack of comparable alternatives we choose it nevertheless. Including $u_0$ in our calculations, the discretized derivative at $r=0$, i.e.,~at $i=1$, immediately fulfills the Neumann b.c.
\begin{equation*}
    \partial_r u(0, t) \approx \frac{1}{2} \left( \frac{u_{2}^{n+1} - u_{0}^{n+1}}{2 \Delta r_1} + \frac{u_{2}^{n} - u_{0}^{n}}{2 \Delta r_1} \right) = 0. 
\end{equation*}
For the governing equation of $u_1$, i.e.~at the boundary, this results in
\begin{equation*}
\begin{split}
        \left( \frac{1}{\Delta t} + \frac{D}{(\Delta r_1)^2} \right) u_1^{n+1} - \frac{D}{(\Delta r_1)^2} u_2^{n+1} + \frac{\gamma}{2} u_1^{n+1} \left( 1 - u_1^{n+1} \right) =  \left( \frac{1}{\Delta t} - \frac{D}{(\Delta r_1)^2} \right) u_1^{n} + \frac{D}{(\Delta r_1)^2} u_2^{n} + \frac{\gamma}{2} u_1^n \left( 1 - u_1^n \right).
    \end{split}
\end{equation*}
The two-dimensional Dirac-Delta distribution whose values on the grid define the initial $(u_i^0)_{i \in \{ 1, ..., I \}}$ is approximated analytically with the following functions, a selection partly inspired by~\cite{DiSchianoCola.2021}:
\begin{align}\label{eq:dirac_delta_approximations}
\begin{split}
  f_1(r) &= \frac{1}{2 \pi r} \cdot \frac{1}{\varepsilon}\; \text{for} \; r < \varepsilon, \; \text{0 otherwise}, \\
  f_2(r) &= \frac{1}{2 \pi r} \cdot \frac{1}{\pi} \cdot \frac{\varepsilon}{r^2 + \varepsilon^2}, \\
  f_3(r) &= \frac{1}{\varepsilon_*}\; \text{for} \;r < 2\varepsilon, \; \text{0 otherwise}, \\
  f_4(r) &= \frac{1}{2 \pi r} \cdot \varepsilon \cdot r^{\varepsilon-1},
\end{split}
\end{align}
where $\varepsilon>0$ is a parameter that directly controls the support only in the first and third approximation. The latter has an additional parameter $\varepsilon_* > 0$ which is chosen such that the integral condition $\int_{\mathbb{R}^2} f_3 (x) \text{d}x = 1$ is fulfilled.
To ensure this condition, and to avoid singularities at $r = 0$, the other approximations $f_1, f_2, f_4$ need to be modified for the numerical implementation:
\begin{align*}
    \begin{split}
        \tilde\delta^2_j(r_i) = \begin{cases}
            c_j(\varepsilon) &\text{if $i = 1, 2$} \\
            f_j(r_i) &\text{if $i \geq 3$ and $r_i < r_\text{lim}$} \\
            0 &\text{if $r_i \geq r_\text{lim}$}
        \end{cases}
    \end{split}
\end{align*}
where $j \in \{ 1, 2, 4 \}$, and $\tilde\delta^2_3 (r_i)= f_3(r_i)$. Details on this modification including the definitions of $c_j$ and $r_\text{lim}$, together with a visualization of these functions, can be found in appendix~\ref{app:dirac_delta_approx_mods}.

\subsection{Biopsy Generation}
The conversion of the solution to the RD-equation to a discrete cell pattern in the plane contains two main tasks: first, the cell density needs to be normalized such that it can be interpreted as a quantity between 0 and 1, and second, an algorithm has to be created which determines where cells are placed according to that quantity.

\begin{figure*}[htb]
\centering 
\includegraphics[width=\columnwidth]{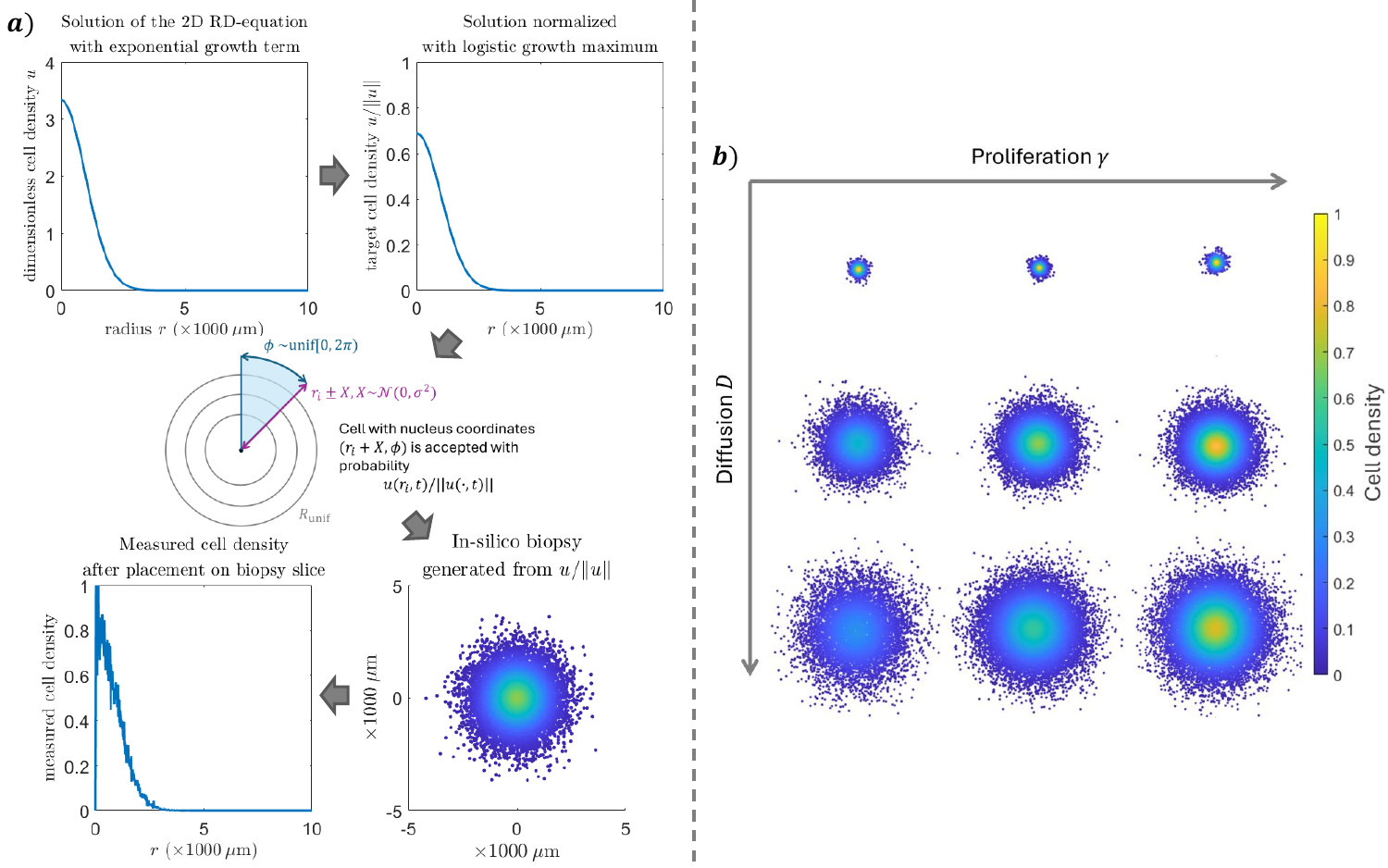} 
\caption{\label{fig:example_biopsies}Cell placement algorithm and exemplary \textit{in silico} biopsies. \textbf{a)} After computing the solutions of the RD equations with the desired values for $D$, $\gamma$ and $t$, $u_\text{exp}$ is normalized with $u_\text{log}$, yielding an $r$-dependent target density for the cells. They are placed on the biopsy following this target density, and the success of the placement can be measured afterwards. \textbf{b)} Exemplary realizations for $D \in \{ 5\:000, 80\:000, 150\:000 \}\frac{\mu\text{m}^2}{\text{day}}$, $\gamma \in \{ 0.3, 0.5, 0.7 \} \frac{1}{\text{day}}$, $t_\text{max}=6$ days. Number of placed cells (row-wise, from top left to bottom right): $1108, 1101, 1104, 8028, 12068, 14849, 10328, 18611, 25415$.}
\end{figure*}

The normalization of $u$ is necessary to make \textit{in silico} biopsies comparable. Concerning the simulation, the comparability is given by well-defined minimum and maximum cell densities: an area contains the minimum cell density if 0 cells are placed in it, and it contains maximum cell density if the placement-algorithm-defined maximum amount of cells is placed in it. These definitions of minimum and maximum density correspond to $u = 0$ and $u = 1$, respectively.
In the case of logistic growth, at least in theory no norm is necessary because the initial condition will be \say{dissolved} infinitely fast by diffusion so that $u = 0$ and $u = 1$ are the extreme values. We introduced the logistic growth term as $\gamma u(1-u)$, which means that growth stagnates at $u = 1$. Since we need to solve the logistic growth RD-equation numerically, $u$-values above 1 are possible due to the Dirac-Delta i.c.~which implies $u>1$ close to the origin for some time. After enough time has passed, diffusion and a regulatory effect of the positive proliferation rate (the growth term $\gamma u(1-u)$ is $<0$ if $u>1$) ensure $u\leq 1$. However, there is no upper bound if the cells grow exponentially, which is the type of growth for which we need the normalization. Apart from $u  \in [0, 1]$, an appropriate normalization also preserves the properties of distinct parameter combinations, eliminating $L^p$-norms as a choice. These types of norms scale $u$ only by diffusion, while it is not possible to derive distinct proliferation rates (as one can see from the analytical solution \eqref{eq:analytical_sol_2D_exp}). A rather pragmatic choice of normalization that mitigates this issue is provided by the RD-equation itself: we compute the solution twice for the same parameter set, once $u_\text{exp}$ for the exponential growth and once $u_\text{log}$ for the logistic growth. Then, we introduce the \say{norm} (which does not fulfill the formal requirements of a norm, hence the quotation marks) 
\begin{equation}\label{eq:unorthodox_norm}
    \| u(\cdot, t) \| := \frac{\| u_\text{exp}(\cdot, t) \|_\infty}{\| u_\text{log} (\cdot, t) \|_\infty} = \frac{\max_r \: u_\text{exp}(r, t)}{\max_r \: u_\text{log} (r, t)},
\end{equation}
which provides a measure of difference between the maximum values of the two kinds of growth. 
In contrast to the $L^p$-norms, this kind of normalization allows to preserve differences in $D$ and $\gamma$. However, it has to be noted that this \say{norm} does not conserve mass equalities between different solutions $u_\text{exp}$. See appendix~\ref{app:rationale_for_normalization} for a demonstration of why this choice of a normalization ensures $u_\text{exp}/\| u \| \in [0, 1]$.

The cell placement algorithm needs to be able to preserve the deterministic concentration profile of $u$ while representing a random process and reducing artifacts in the spatial statistic to a minimum. 

The polar geometry from the PDE solver is used again for the placement, by iterating through circles around the origin. For cost efficiency, their radii can be defined by the original coarse $R_\text{unif}$. In the $i$th iteration, $N_i$ cell positions $(r_i + X, \phi)$ are generated by selecting $\phi \in [0, 2\pi)$ from a uniform distribution, and adding Gaussian noise by sampling $X$ from $\mathcal{N} (0, \sigma^2)$, where $\sigma > 0$ is equal for all iterations. $N_i$ is determined from the circumference of the circle with radius $r_i$, i.e.,~if $s$ is the cell size, $c_i$ the circumference, then $N_i = c_i / s$. If desired, one can perform a collision check before placing the cells on the plane, but our tests show that the continuous density profile is better preserved if one ceases to do so, see appendix~\ref{app:collision_cell_size} for more information. Eventually, the new cells are filtered with the normalized concentration: each of the remaining points has a probability of $u(r_i, t_\text{max})/ \| u(\cdot,t_\text{max}) \|$ to make it to the plane. An illustration of the placement algorithm and exemplary realizations are shown in Fig.~\ref{fig:example_biopsies}.

\subsection{Validation Procedure}
The \textit{in silico} biopsies are generated using various combinations of input parameters $(D, \gamma)$,
and the SSA method (in the version that uses the convex hull of the biopsy point pattern as input data) then tries to retrieve these parameters, generating output parameters $(\hat{D}, \hat{\gamma})$ which are not direct approximations of their input counterparts. Instead they follow certain conversion laws which one can determine by considering the relationship between in- and output parameters and the mechanisms of the method (for instance, the cell placement algorithm, or the norm).

First experiments revealed that $\hat{D}$ and $\hat{\gamma}$ need to be converted to $D$ and $\gamma$ sequentially, not simultaneously. In the case of exponential growth, which is the type of growth that is mainly considered in this work, $\hat{D}$ is linked to $D$ via an inverse-quadratic relationship, i.e.,~the conversion law reads as
\begin{equation}\label{eq:conversion_formula_D}
    D \approx f(\hat{D}) = a_f \sqrt{\hat{D}} + b_f \hat{D} + c_f,
\end{equation}
where $a_f, b_f, c_f$ are to be determined from the data. Most likely, the reason for the inverse-quadratic nature is a data-handling-\say{artifact}: the point pattern is handed to the 2PCF in cartesian coordinates, which are then handled in polar coordinates by the conversion formula.

$\hat{\gamma}$ is highly dependent on $D$ (or $\hat{D}$), as a larger diffusion coefficient allows for a more expressed proliferation (see Fig.~\ref{fig:example_biopsies} b). This of course is an effect of the norm which we had to introduce, but low and high proliferation rates are still distinguishable, their scale just depends on $D$. Hence, the conversion law reads
\begin{equation}
\label{eq:alt_conversion_formula_gamma}
    \gamma \approx g(\hat{\gamma}, \hat{D}) = a_g \hat{\gamma} + b_g f(\hat{D}) + c_g,
\end{equation}
again with $a_g, b_g, c_g$ from the data. An important difference to \eqref{eq:conversion_formula_D} is that $a_g, b_g, c_g$ are not the same for all $\hat{\gamma}$ one wishes to convert, but they depend on $f(\hat{D})$ (as in the validation we pretend that we do not know the true $D$). We determine $a_g, b_g, c_g$ group-wise by dividing the pairs $(\hat{\gamma}, \hat{D})$ into several groups with relatively similar $f(\hat{D})$-values (where \say{similar} can have different definitions depending on the numbers of biopsies tested, and it has to be ensured that each group contains at least three pairs $(\hat{\gamma}, \hat{D})$ to be able to determine $a_g, b_g, c_g$). While the data suggests a quadratic relationship between $f(\hat{D})$ and $a_g, b_g, c_g$, the conversions were more successful using piecewise-linear functions connecting the parameter values of the different groups.

We will also see that there are instances where $D$ and $\gamma$ cannot be retrieved separately, only the expression $D/\gamma$ can be estimated using a combination of \eqref{eq:conversion_formula_D} and \eqref{eq:alt_conversion_formula_gamma}:
\begin{equation}\label{eq:conversion_D_over_gamma}
    \frac{D}{\gamma} \approx  h(\hat{D}, \hat{\gamma}) = \frac{a_f\sqrt{\hat{D}} + b_f\hat{D} + c_f}{a_g\hat{\gamma} + e_h \sqrt{\hat{D}} + f_h\hat{D} + g_h},
\end{equation}
where in theory $e_h = a_fb_g$, $f_h = b_fb_g$, $g_h = b_gc_f + c_g$, but in practice all parameters of \eqref{eq:conversion_D_over_gamma} need to be re-estimated, as $D$ and $\gamma$ cannot be obtained independently. 
The biomarker $\sqrt{D/\gamma}$ is then estimated by subsequent square root extracting. For $D \cdot \gamma$ no such combined formula exists due to non-unique parameter relationships in the solution of \eqref{eq:2D_polar_RD_exp_growth}, see appendix \ref{app:retrieval_of_Dtimesgamma} for details.

\section{Experimental Setup}

A common problem in mathematical oncology is that some studies randomly pick parameter values from literature and thereby mix different types of cancer, leading to inconclusive results for the clinical application~(\cite{Yankeelov.2013}). It should be emphasized at this point why this does not apply here: this work does not aim to reflect the exact biological conditions, but merely assesses the SSA method under known input conditions. The clinical relevance is covered by~\cite{Pasetto.2024} and~\cite{PirminsNewPaper}. Still, it is desirable to use biologically plausible parameter ranges to avoid unrealistic model phenomena.

There are two types of parameters that need to be determined for the solution of the PDE and the creation of the biopsies: numerical and biological parameters. While the numerical parameters (such as grid node distance or time step size) are mainly derived from the biological parameters and/or set experimentally by testing which values lead to the best convergence, the values of the biological parameters are based on values from literature and reasoning. All parameter values that are used in this work are summarized in Tab.~\ref{tab:parameter_vals}, and their choice is justified in the following.

\begin{table*}[!htb]
\centering
\begin{tabular}{|c | c | c | c | c |}
\hline
Parameter & Unit & Values & Optimum  & Source \\
& & & (if applicable) & \\
\hline\hline
Diffusion coefficient $D$ & $\frac{\mu\text{m}^2}{\text{day}}$ & $[5\:000, 150\:000]$ & - & Swanson et al.~2008, \\ 
 & & or $[5\:000, 20\:000]$ & & and depending on $t_\text{max}$ \\
\hline
Proliferation rate $\gamma$ & $\frac{1}{\text{day}}$ & $[0.3, 0.7]$ & - & - \\
\hline
Cell radius $s$ & $\mu \text{m}$ & 11 & - & \cite{Khetan.2019} \\
\hline
Placement std. $\sigma$ & $\mu \text{m}$ & 20 & - & - \\
\hline 
Uniform grid distance $\Delta r$ & $\mu \text{m}$ & $1\:000, 500, 250, ...$ & 15.625 & experimental \\
\hline
Maximum radius $r_\text{max}$ & $\mu \text{m}$ & $10\:000$ & - & experimental \\
\hline
Time step size $\Delta t$ & day & depending on~\eqref{eq:CFL} & depending on~\eqref{eq:CFL} & - \\  
\hline
Simulation end time $t_\text{max}$ & day & $\mathbb{N}$ & 6 & experimental \\
\hline
PSD \& 2PCF argument $t_\text{max}$ & day & $\{ 1, ..., 10 \}$ & - & - \\
\hline
$\tilde{\delta}^2$-parameter $\varepsilon$ & $\mu \text{m}$ & $\{ 625, 312.5, 156.25,$ & 78.125 & depending on $r_\text{max}$ \\
 & & $ 78.125 \}$ & & \\
\hline
\end{tabular}
\caption{Parameter ranges and/or optimum parameter values used in this work.}
\label{tab:parameter_vals}
\end{table*}

Three crucial biological parameters are the diffusion coefficient, the proliferation rate and the cell size. Swanson et al.~find in their work on the RD-model of glioma cells (the cells causing glioblastoma, a brain tumor) a mean diffusion coefficient of $D \approx 120\:000 \:\frac{\mu \text{m}^2}{\text{day}}$ by considering MRI data~(\cite{Swanson.2008}). Glioblastoma is a cancer known for its extremely aggressive spread, hence this value can be assumed as an upper limit for the choice of $D$ used in the simulations. Assuming exponential growth (i.e.,~$u(t) = u_0 e^{\gamma t}$) and a cell density doubling time of one day (which is extremely fast and can be taken as a lower limit for doubling time), we can make the following estimation of an upper bound for $\gamma$:
\begin{align*}
    u_0 e^{\gamma (t+1)} &= 2u_0 e^{\gamma t} \\
    \Leftrightarrow \gamma &= \text{ln}(2) \approx 0.7,
\end{align*}
where the $t$ and $t+1$ in the exponents come from the premise that at day $t+1$, we have double the amount of cells we had at day $t$.
To avoid zero as lower bound for these parameters (because no spread/growth would not be considered cancer), we choose $D \geq 5\:000 \:\frac{\mu \text{m}^2}{\text{day}}$ and $\gamma \geq 0.3 \:\frac{1}{\text{day}}$.

Cell sizes vary following a log-normal distribution as suggested by observations in breast cancer cells~(\cite{Khetan.2019}). In their study, Jawahar Khetan et al.~find that the approximate distribution of the cell radius $\theta$ is log-normally distributed with mean $11 \mu \text{m}$ and standard deviation $0.2 \mu \text{m}$. As mentioned, in its current version the biopsy generating algorithm needs cell sizes only to obtain an estimate for the number of cells that are to be placed in each iteration, not an exact size distribution. We will thus only use this mean and fix $s = 11 \mu \text{m}$. The resulting cell diameter of $22 \mu \text{m}$ is used as reference point for the standard deviation in cell placement, and we set $\sigma = 20 \mu \text{m}$.

The numerical parameter $r_\text{max} = 10\:000 \:\mu \text{m}$ is chosen from evaluations of the analytical solution~\eqref{eq:analytical_sol_2D_exp} with the mentioned parameter ranges for $D$ and $\gamma$: $u$ is equal to zero (at least at numerical precision) at this point even for the fastest spreading, most proliferative combination $(D, \gamma) = (150\:000, 0.7)$.

The spatial and temporal step sizes that will be evaluated in terms of the convergence of the solution obtained with them start at more or less arbitrary values and are then simultaneously refined. The relationship between space and time step size during this refinement should follow a Courant-Friedrichs-Lewy (CFL)-like condition as a necessary prerequisite of convergence (see~\cite{Courant.1928} for the original condition, and~\cite{Trefethen.1996} for some details on parabolic and nonlinear PDEs). For the reaction-diffusion equation, the condition looks like
\begin{equation}\label{eq:CFL}
    \frac{\Delta t}{(\Delta r)^2} \leq a |D|,
\end{equation}
where $a > 0$ scales the upper bound of the Courant number (the term on the left hand side). The smaller the stability region of a numerical method, the smaller $a$, and hence the stricter the condition. If the reaction term is linear, $a$ can be chosen arbitrarily for the Crank-Nicolson scheme in one dimension. But as we are not only in more-dimensional space, but also have the issue of a nonlinear reaction term, we use the very conservative choice of $a = \frac{1}{2}$, which is even convergent for the explicit Euler method~(\cite{Trefethen.1996}). 

The scaling parameter $\varepsilon$ of the Dirac-Delta approximations is tested for several values depending on the size of the spatial domain: $\{r_\text{max}\cdot2^{-4}, r_\text{max}\cdot2^{-5}, r_\text{max}\cdot2^{-6}, r_\text{max}\cdot2^{-7} \} = \{625, 312.5, 156.25, 78.125 \}$.

$t_\text{max}$ has an ambiguous definition in the context of the validation. The PSD and 2PCF with which the curve fit is performed need a time variable as input argument, which is taken to be some fixed value to allow for better inter-patient comparability. In the validation, we aim to find out about how tumor age influences the quality of parameter estimation, hence we need some variation in $t_\text{max}$. We have two possibilities to implement this variation: either we select some time constant for PSD and 2PCF that is valid for all experiments and construct the biopsies for each parameter combination $(D, \gamma)$ separately for every $t_\text{max}$ (such that we really have $(D, \gamma, t_\text{max})$ as parameter combinations for the \textit{in silico} biopsies), or we use this fixed time constant to construct one biopsy per $(D, \gamma)$-combination and vary the time argument of PSD and 2PCF. For most of the conducted experiments, we decide for the latter because it allows to use the same simulation multiple times, saving computational cost.

Experiments have shown that the fixed $t_\text{max}$ that is used for all simulations should best be set to $t_\text{max} = 6$ days because then it is as small as possible to minimize the computational cost of the simulation, yet as large as necessary to allow diffusion to progress enough so that the solution $u$ fulfills $u(r, t) \leq 1$ (a numerical phenomenon coming from the Dirac-Delta i.c.). This is the $t_\text{max}$-value that is also used for the assessment of the numerical solver. However, we also use $t_\text{max} = 100$ days for some simulations to obtain insights into the validity of our results in a long-term setting. In this scenario, we restrict the upper limit of values for $D$ to $20\:000 \:\frac{\mu \text{m}^2}{\text{day}}$ to make long observation times computationally feasible.  

The $t_\text{max}$ which is used for PSD and 2PCF is set to integers between 1 and 10 days to have an overview in the short term, and in an additional experiment to $\{ 50, 100, 200 \}$ days for an idea of the long-term behavior. 

The choice of relatively short time periods in both use cases of $t_\text{max}$ is justified by clinical practice: Hutchinson and Grimm, who work on biopsy-based biomarkers as well, used pretreatment and on-treatment biopsy pairs from a wide range of cancers which were taken 15-113 days apart from each other~(\cite{Hutchinson.2022}), i.e.,~a comparable order of magnitude to our range of 6-100 days and 1-200 days, respectively.

The \textit{in silico} biopsies created with these parameters are used for two types of validation experiments: in both types, we create biopsies with different $(D, \gamma)$-combinations (see appendix~\ref{app:D_gamma_combis}) and apply the SSA method with different $t_\text{max}$ to them to receive corresponding $(\hat{D}, \hat{\gamma})$-pairs. In the first experiment we know which parameters were estimated with the same PSD- and 2PCF-$t_\text{max}$, even though we do not know the value of $t_\text{max}$. We call the groups of $(\hat{D}, \hat{\gamma})$-pairs that were estimated with the same $t_\text{max}$ \say{$t$-cohort}, and we use \eqref{eq:conversion_formula_D} and \eqref{eq:alt_conversion_formula_gamma} to retrieve estimates of the original input parameters $(D, \gamma)$ for every $t$-cohort separately. In the second experiment, we do not use any information about $t_\text{max}$, i.e.,~one could say we mix the $t$-cohorts. With the lack of information about the same point in time, we cannot use the separate conversions anymore due to the time normalization and estimate only the biomarker $\sqrt{D/\gamma}$ using \eqref{eq:conversion_D_over_gamma}.

\begin{figure*}[htb]
\centering 
\includegraphics[width=\columnwidth]{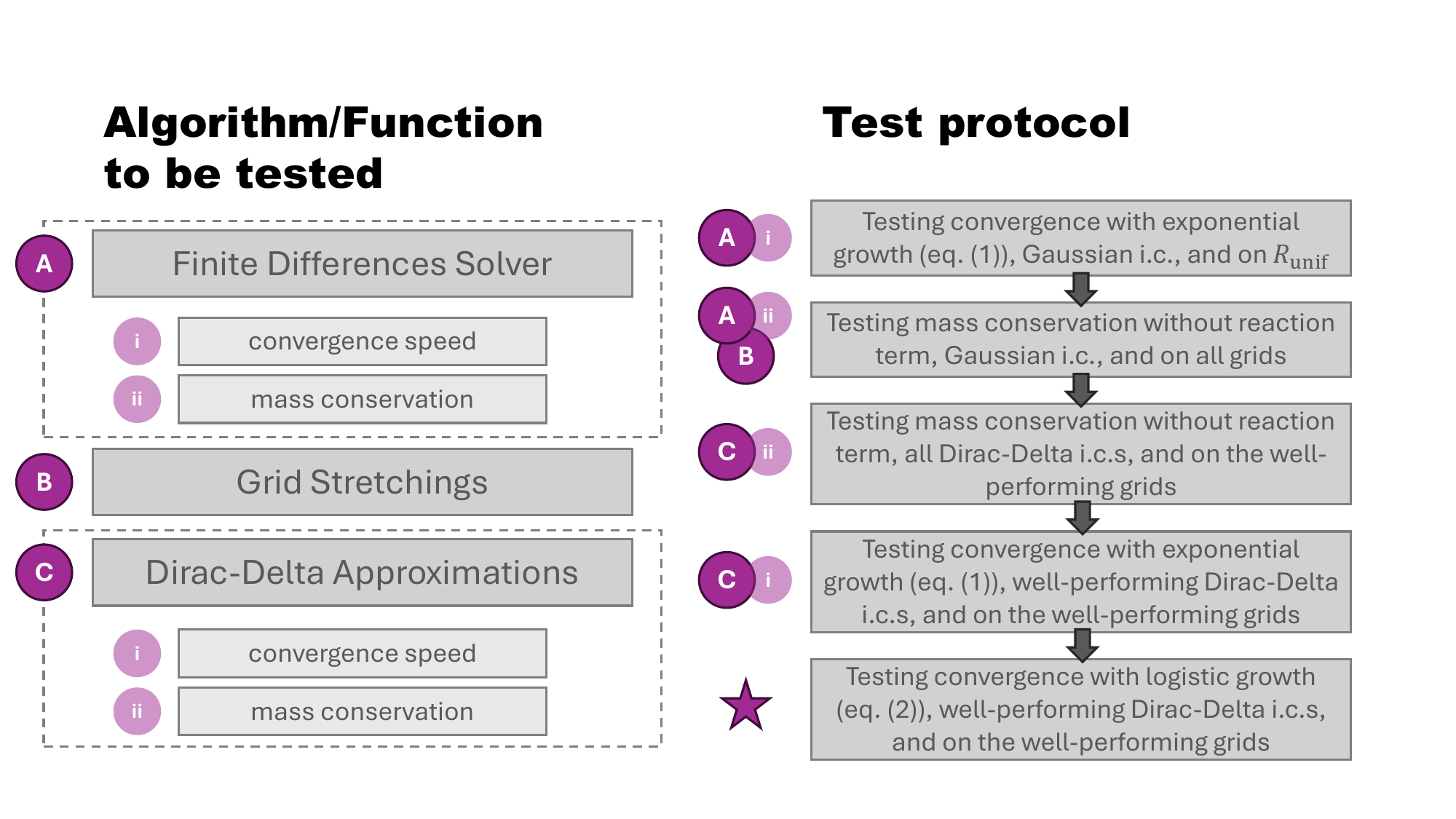} 
\caption{\label{fig:solver_testing_procedure}Overview of the tests that are performed with the PDE solver.}
\end{figure*}

\section{Results}

\subsection{PDE Solver}
Testing the solver and the Dirac-Delta approximation are distinct tasks, which are visually structured in Fig.~\ref{fig:solver_testing_procedure}. First, we test the solver with the most basic options: on $R_\text{unif}$, using an unproblematic, continuous i.c.~and exponential growth term, and we compare the result to the analytical solution (A i, A ii, B). Next, we use the Dirac-Delta approximations and the grid stretchings and test for mass conservation, because the steep drop of the i.c.~close to the origin is the critical point where mass might be added or lost (C ii). We then add exponential growth and check whether well-conserving grids converge to the analytical solution (C i). Only in the last step, we use the logistic growth term and assess convergence speed ($\star$).

As a continuous i.c.~we choose the aforementioned Gaussian curve
\begin{equation}\label{eq:gaussian_initial_condition}
    u(r, 0) = \frac{1}{\pi} e^{-r^2}.
\end{equation}
The analytical solution of~\eqref{eq:2D_polar_RD_exp_growth} with this i.c.~is given by
\begin{equation*}
    u(r, t) = \frac{1/ \pi}{1 + 4 Dt} e^{-\frac{r^2}{1+4Dt} + \gamma t}.
\end{equation*}
\begin{figure*}[htb!]
\centering 
\includegraphics[width=\columnwidth]{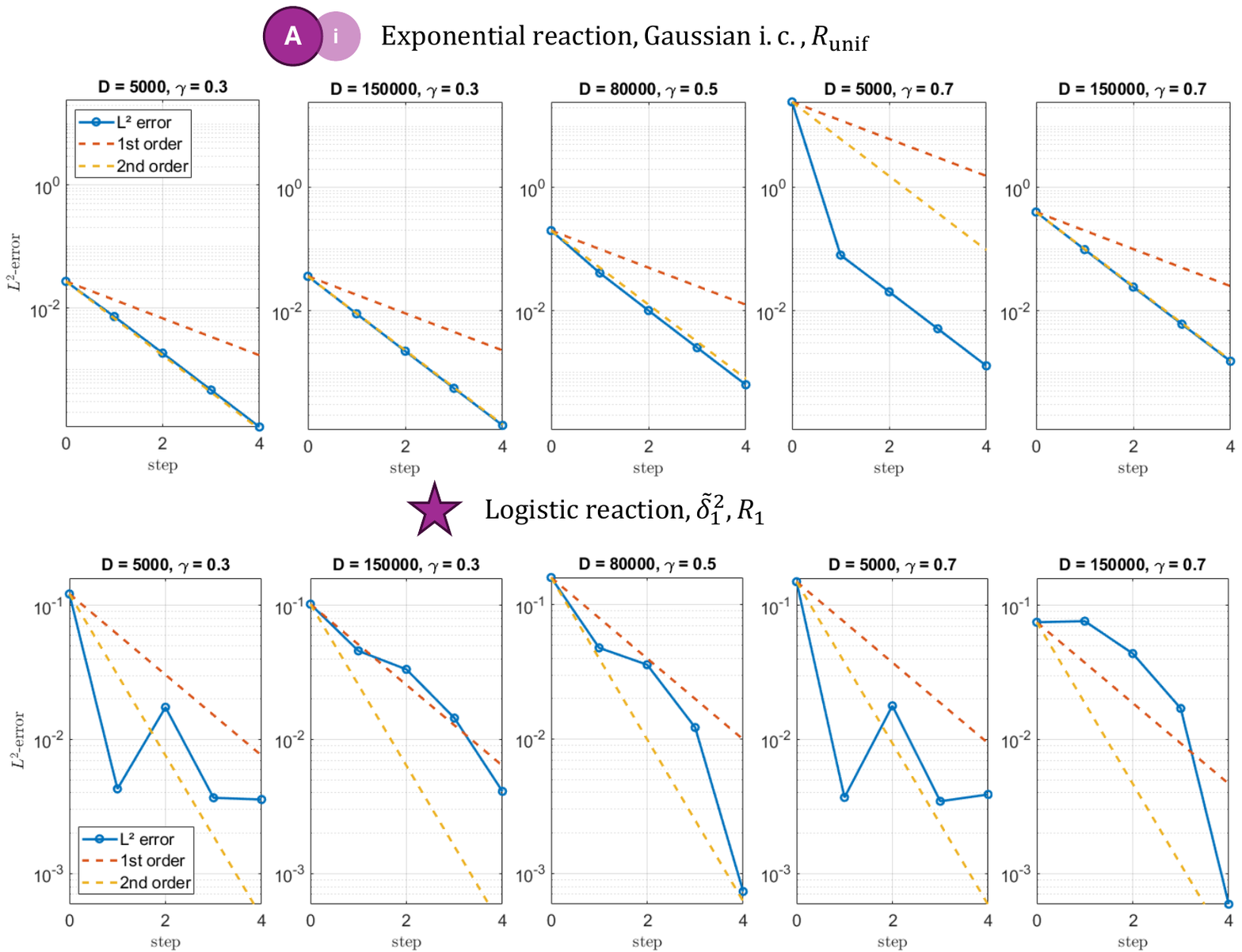} 
\caption{\label{fig:test_Ai_and_star}Convergence (measured via $L^2$-error) of the numerical solutions for the four extreme $(D, \gamma)$-combinations and a combination from the center of the intervals specified in Tab.~\ref{tab:parameter_vals} ($D = 80\:000 \frac{\mu\text{m}^2}{\text{day}}$, $\gamma = 0.5 \frac{1}{\text{day}}$). In the horizontal axis, ``step'' is a measure for the spatial and temporal grid refinement: starting with the initial spatial step width $\Delta r_0 := 1\:000 \:\mu\text{m}$, we define $\Delta r(\text{step}) = \Delta r_0 \cdot 2^{-\text{step}}$, and, in accordance with \eqref{eq:CFL}, $\Delta t(\text{step}) = \frac{1}{2}\Delta r(\text{step})^2D$.
\textbf{Top row}: Test for A i, i.e., RD-equation with exponential reaction term \eqref{eq:2D_polar_RD_exp_growth} and Gaussian i.c.~\eqref{eq:gaussian_initial_condition} on $R_\text{unif}$. 
\textbf{Bottom row}: Test for $\star$, i.e., RD-equation with logistic reaction term \eqref{eq:2D_polar_RD_log_growth} and the i.c.~$\tilde\delta^2_1$ on grid $R_1$.
}
\end{figure*}
\begin{figure*}[htb!]
\centering 
\includegraphics[width=\columnwidth]{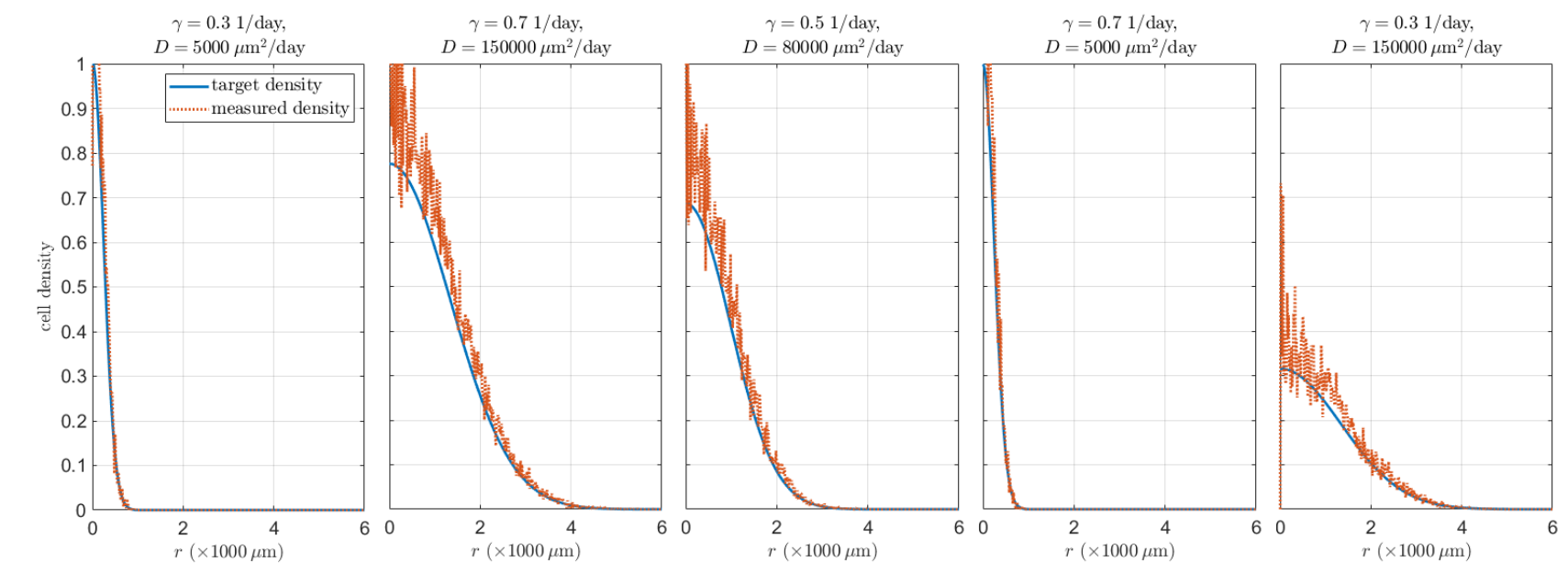}
\caption{\label{fig:measured_vs_desired_density_overlapping_ok}Target versus measured cell density in the \textit{in silico} biopsy for different parameter combinations, this time without controlling for overlapping cells. In this context, target density means the density computed by $\| u(r, t_\text{max}) \|$, and measured density refers to the density of digital cells in the biopsy. The $(D, \gamma)$-combinations are the same as in the convergence analysis.} 
\end{figure*}
The result of the initial test A i is shown in the top row of Fig.~\ref{fig:test_Ai_and_star}, and demonstrates clear 2nd order convergence in $L^2$-error (the way we defined it above), as it is expected from a Crank-Nicolson scheme. The mass conservation test A ii, B reveals that for the Gaussian i.c., all grids do reasonably well (see Fig.~\ref{fig:test_ABii} in appendix~\ref{app:solver_tests_additional_figs}). Switching to the Dirac-Delta approximations as i.c.s in test C ii, we find that the first and the third Dirac-Delta approximations $\tilde\delta^2_1, \tilde\delta^2_3$ are the best choices for mass conservation, provided we use a grid different from $R_\text{unif}$, and $\varepsilon$ is small enough (see Fig.~\ref{fig:test_Cii} in appendix~\ref{app:solver_tests_additional_figs}). Convergence test C i provides another argument in favor of the first Dirac-Delta approximation $\tilde\delta^2_1$, with the best results on the grid stretchings $R_2$, $R_3$ and $R_4$ (see Fig.s~\ref{fig:test_Ci_Runif}-\ref{fig:test_Ci_R4} in appendix~\ref{app:solver_tests_additional_figs}). Grid stretchings are supposed to be smooth functions to prevent numerical instability and to maintain low errors~(\cite{Fornberg.2009}), hence it is no surprise that the two smooth stretchings $R_3$ and $R_4$ are among the best-performing options. 
It has to be noted that in the case of logistic growth, the solver only reliably works with $R_\text{unif}$ and $R_1$ because of several complications: for instance, in the early time steps the Jacobi-matrix that is used by the numerical solver has been observed to be more easily ill-conditioned with non-linear grid transformations such as $R_2$-$R_4$ than with linear ones, i.e.,~$R_1$. Also, ghost point distances at the boundaries are undefined in the case of non-uniform node distances and therefore potentially wrong. Issues like this are more prominent for the logistic than for the exponential growth because error protrusion is quadratic instead of linear. This is why eventually, grid stretching $R_1$ is chosen, even though it does not expose as fast -- but still reliable -- convergence together with $\tilde\delta^2_1$.
The final test $\star$ for convergence speed with this set-up and logistic growth shows that convergence is decent, see the bottom row of Fig.~\ref{fig:test_Ai_and_star}.

\subsection{Biopsy Generation}

With $R_1$ and $\tilde\delta^2_1$, we continue to compute $\| u(r, t_\text{max}) \|$ for the parameter combinations $(D, \gamma)$ listed in appendix~\ref{app:D_gamma_combis}, where we use the analytical solution for $u_\text{exp}$ and the numerical one for $u_\text{log}$ (in the context of creating the biopsies, $t_\text{max}$ is the simulation parameter, i.e.,~$t_\text{max}=6$ days or 100 days). This means that we only use the PDE solver to obtain a single value for each $(D, \gamma)$-pair, namely $\text{max}_r \: u_\text{log}(r, t_\text{max})$. In total, we compute \textit{in silico} biopsies for 38 parameter combinations in the short-term experiments, and for 28 parameter combinations in the long-term experiments.

\begin{figure*}[htb!]
\centering 
\includegraphics[width=\columnwidth]{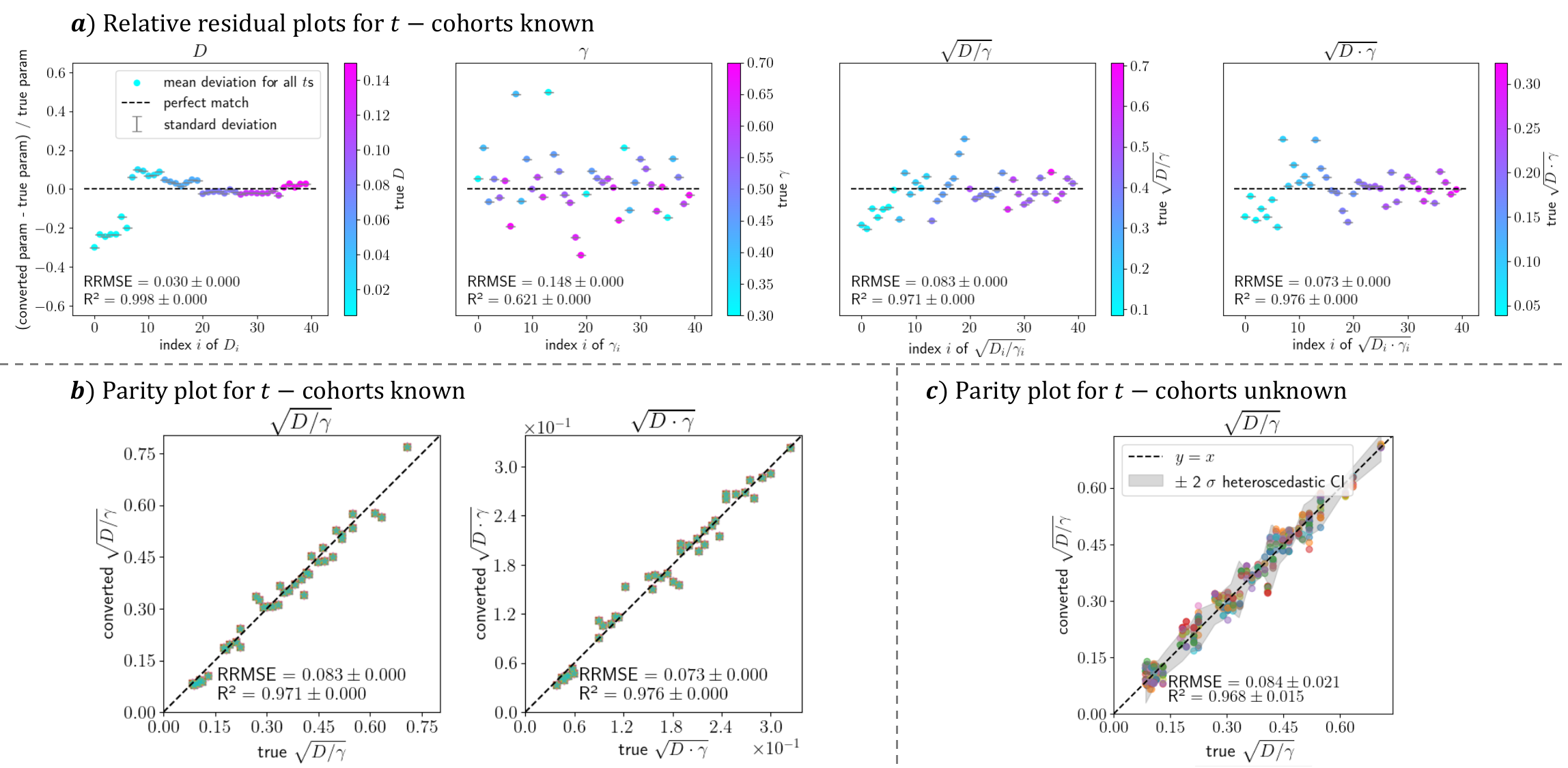} 
\caption{\label{fig:parity_exp_growth}Comparison of input parameter values $D, \gamma$ and converted output parameters. 
\textbf{a)} Relative residual plots if the $t$-cohorts are known. We compute the conversions $f(\hat{D})$, $g(\hat{\gamma}, \hat{D})$ using \eqref{eq:conversion_formula_D} and \eqref{eq:alt_conversion_formula_gamma}, and compute the output biomarkers as $\sqrt{f(\hat{D})/g(\hat{\gamma}, \hat{D})}$ and $\sqrt{f(\hat{D}) \cdot g(\hat{\gamma}, \hat{D})}$. The (hardly visible) error bars illustrate the standard deviation between different values of $t_\text{max}$.
\textbf{b)} To account for visual biases in a), we show the parity plots for the biomarkers as an illustration.
\textbf{c)} Parity plot of input parameter ratio $\sqrt{D/\gamma}$ and converted output parameter ratio $\sqrt{h(\hat{D},\hat{\gamma})}$ in the case of unknown $t$-cohorts.
Each color stands for one round of sample drawing.
}
\end{figure*}

Placing the cells according to the above-described algorithm leads to volume-filling distributions shown in Fig.~\ref{fig:measured_vs_desired_density_overlapping_ok} (for $t_\text{max} = 6$ days; the data for $t_\text{max} = 100$ days is not shown but comparable). Close to the origin the match is not perfect but there is a home-made reason for this: the smaller the area that can be covered by the cells, the more overlaps that are not respected by the measure of cell-covered area, which adds the areas of every cell without taking into account that some cells essentially share the same space. In the case of an algorithm that avoids overlapping cells, this measure is more accurate, but the strict rules on placement may lead to an under-filled biopsy, see appendix~\ref{app:collision_cell_size}.

\subsection{Validation}

The \textit{in silico} biopsies are used as input for the SSA method, which is able to estimate $\hat{D}$ and $\hat{\gamma}$ from the point pattern by fitting the pattern's 2PCF and PSD with $\text{R}^2$-values above 0.9 in about $90\%$ of the tested biopsies. Now, we use the conversion formulae \eqref{eq:conversion_formula_D}, \eqref{eq:alt_conversion_formula_gamma} and \eqref{eq:conversion_D_over_gamma} 
to retrieve $D$ and $\gamma$.

We start with the results for the short-term observations. In the scenario with the known $t$-cohorts, we can convert and retrieve $D$ and $\gamma$ separately and receive the results shown in Fig.~\ref{fig:parity_exp_growth} a) and b). The errors for the single parameters $D$ and $\gamma$ range between $3.0\%$ and $14.8\%$ (RRMSE), and 0.621 and 0.998 ($\text{R}^2$), where especially the recovery of $D$ performs well. The standard deviations are negligible, so they are omitted here. In total, for the biomarker $\sqrt{D/\gamma}$ we obtain an RRMSE of $8.3\%$ and an $\text{R}^2$ of 0.971, and for $\sqrt{D\cdot \gamma}$ it is $7.3\%$ and $0.976$, respectively. It should also be mentioned that Fig.~\ref{fig:parity_exp_growth} a) demonstrates how robust the conversions \eqref{eq:conversion_formula_D} and \eqref{eq:alt_conversion_formula_gamma} are under different values for $t_\text{max}$: if some biopsies were created with the same $(D, \gamma)$ but the SSA method was applied with different $t_\text{max}$, then $(f(\hat{D}), g(\hat{\gamma}, \hat{D}))$ will be the same for all of these biopsies, even though the $(\hat{D}, \hat{\gamma})$-pairs that result from the SSA differ. Refer to appendix~\ref{app:validation_details} for more details.

The more realistic setting in which we do not know which biopsies' parameters were estimated with the same PSD- and 2PCF-$t_\text{max}$ leads to an average $\sqrt{D/\gamma}$-RRMSE of $8.4\% \pm 2.1\%$ and corresponding $\text{R}^2$ of $0.968 \pm 0.015$,
where 15 rounds of 30 samples were tested (i.e., in each round 30 $(\hat{D}, \hat{\gamma})$-pairs from unknown $t_\text{max}$ were randomly selected and a curve fit with \eqref{eq:conversion_D_over_gamma} was performed with them). Figure~\ref{fig:parity_exp_growth} c) shows the parity plot for all 15 rounds, including a heteroscedatic confidence interval of two standard deviations, demonstrating that 
$\sqrt{D/\gamma}$ could be retrieved very precisely.

\begin{figure*}[htb]
\centering 
\includegraphics[width=\columnwidth]{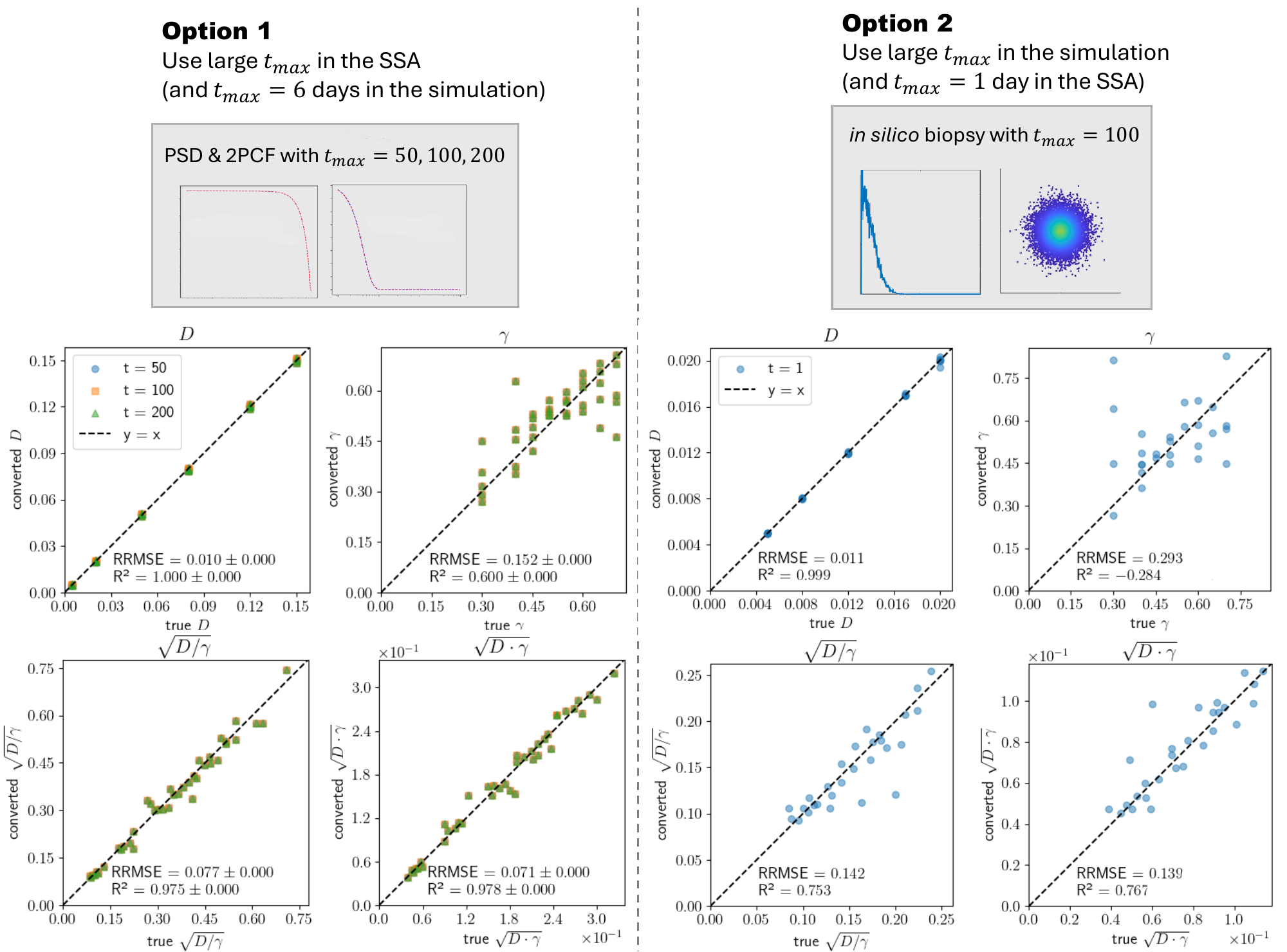} 
\caption{\label{fig:large_t_parity} Parity plots of the long-term experiments ($t$-cohorts known). \textbf{Left}: Large time scale in the SSA functions (PSD, 2PCF), short time scale in the \textit{in silico} biopsies. The markers for all tested $t_\text{max}$ coincide as a result of the robustness of the converted values against the input time argument. \textbf{Right}: Vice versa, but with only one $t$-cohort.}
\end{figure*}

We conclude the section with a summary of the results from the long-term observations. When large $t_\text{max}$-values are used as input parameters of the SSA-functions, we observe RRMSEs of $1.0\%$ ($D$), $15.2\%$ ($\gamma$), $7.7\%$ ($\sqrt{D/\gamma}$) and $7.1\%$ ($\sqrt{D\cdot\gamma}$), and $\text{R}^2$-values of $1.000$ ($D$), $0.600$ ($\gamma$), $0.975$ ($\sqrt{D/\gamma}$) and $0.978$ ($\sqrt{D\cdot\gamma}$), in the case of known $t$-cohorts. Again, the conversions are robust under different $t_\text{max}$-values which is visible in the negligible standard deviations for the different $t_\text{max}$. See the left column of Fig.~\ref{fig:large_t_parity} for the parity plots. In the case of unknown $t$-cohorts we obtain an $\sqrt{D/\gamma}$-RRMSE of $7.5\% \pm 2.4\%$ 
and an $\text{R}^2$ of $0.975\pm 0.015$ 
(data not shown). 

The other option, where $t_\text{max}=100$ days is used for the simulation, provides similar results only for $D$: the RRMSEs are given by $1.1\%$ ($D$), $29.3\%$ ($\gamma$), $14.2\%$ ($\sqrt{D/\gamma}$) and $13.9\%$ ($\sqrt{D\cdot \gamma}$), and $\text{R}^2$-values of $0.999$ ($D$), $-0.284$ ($\gamma$), $0.753$ ($\sqrt{D/\gamma}$) and $0.767$ ($\sqrt{D\cdot \gamma}$), without denoting standard deviations because only one $t$-cohort is tested (right column of Fig.~\ref{fig:large_t_parity}). The negative $\text{R}^2$-value of $\gamma$ is most likely due to a too coarse $f(\hat{D})$-dependent grouping of the parameter pairs, which could not be refined as otherwise there would not have been enough pairs in each group to determine the parameters of \eqref{eq:alt_conversion_formula_gamma}. The well-retrievable $D$ is able to balance the effect for the biomarkers.

\section{Discussion}
In this work we have demonstrated a method to solve two-dimensional RD-equations with Dirac-Delta i.c.~under the assumption of radial symmetry, an algorithm that creates discrete cell patterns from a continuous density, and we were able to show that the previously published SSA method estimates the ratio and, if some information about the tumor age is available, the product of diffusion coefficient and proliferation rate from a single biopsy within a certain tolerance.

While our proposed finite differences solver did not consistently reach 2nd order convergence for logistic growth and the approximated Dirac-Delta i.c., we still observe convergence sufficient for the purposes of this work, as well as mass conservation for all parameter combinations while avoiding singularity or stiffness issues. Future works on solvers for \eqref{eq:2D_polar_RD_exp_growth} and \eqref{eq:2D_polar_RD_log_growth} with discontinuous, complicated i.c.s should focus on obtaining similar convergence properties as one observes with continuous i.c.s.

The cell placement algorithm that is used for the \say{reverse coarse graining} is able to follow the continuous cell density governed by the RD-equation as long as cell sizes are ignored. This issue could be mitigated by allowing for compressible cells: in two dimensions, the maximum packing density of equal spheres is around $91\%$~(\cite{Fejes.1942}). While the spheres that constitute the cells are not necessarily of the same size, it is clear that even in the case of irregular sphere packing not the whole domain can be filled with incompressible spheres, making $100\%$ area coverage unattainable. 

The algorithm has another flaw which did not necessarily impact the results presented here but should be the focus of future works: the strategy by which the radial component of the cells' coordinates is selected is likely to introduce artifacts to the spatial statistics of the SSA method. Currently, Gaussian noise with a fixed standard deviation is added to a fixed radius -- a highly systematic approach without enough room for randomness. This could be an issue because the SSA method computes the 2PCF of the biopsy which is then transformed into the PSD, and the PSDs from \textit{in silico} biopsies differ qualitatively from PSDs that were derived from real biopsies: the \textit{in silico} PSDs strongly oscillate, while the real-biopsy-PSDs are almost monotonous functions. The subsequent curve fit that estimates $\hat{D}$ and $\hat{\gamma}$ might be impacted by these oscillations. The validation results presented here do not seem to be influenced substantially by this as they are still of acceptable quality, but for other experimental settings this issue should be addressed. One possibility to do this would be by avoiding the iteration through the radii and instead sample a radius from a $u$-shaped distribution for each cell individually.

An instance where the flawed placement algorithm might play a major role is the switch of growth terms: this work presented the validation of the SSA method with normalized exponential growth, whereas in many scenarios logistic growth may be more appropriate (e.g.~damped proliferation due to lack of resources). Appendix~\ref{app:logistic_growth_validation} provides the results of the validation performed with logistic growth, which failed completely. Apart from the faulty placement algorithm, it is also possible that the SSA method, which uses closed-form expressions for 2PCF and PSD in the curve fit that are derived from the exponential growth RD-equation, is not perfectly suitable for tumors that developed exclusively under the logistic growth law. It will be interesting to see whether the method performs better under a logistic-growth-derived PSD and 2PCF.

If we however use the \say{combined} scenario of the logistic-growth-normalized exponential growth -- as we did here -- we find RRMSEs and $\text{R}^2$-values that lie close to or even well within the acceptable range, respectively. The RRMSE is not an error measure that is commonly used in medical research, but publications that consult it set values of 20 to 30\% as threshold for clinically relevant results~(\cite{Uster.2021, Schouwenburg.2025}).
The $\sqrt{D/\gamma}$- and $\sqrt{D\cdot\gamma}$-RRMSEs of both short-term experiments (the one with knowledge about shared $t_\text{max}$, and in the case of $\sqrt{D/\gamma}$ also the one without this knowledge), as well as the biomarker RRMSEs of the long-term experiment with the large $t_\text{max}$ as SSA input, fulfill the stricter of these limits -- a promising result given the flaws of the biopsy generation procedure. Regarding $\text{R}^2$, we can be similarly optimistic: its average values for $\sqrt{D/\gamma}$ and $\sqrt{D\cdot\gamma}$ range between 0.968 and 0.978 for the mentioned experiments, signaling that the variability in parameter ratio can be very well recovered overall. Only in the experiment with large $t_\text{max}$-values in the simulation, the quality of the parameter recovery drops to an RRMSE of about $14\%$ and an $\text{R}^2$ around $0.75$ for both biomarkers. There are however possible explanations for this that have nothing to do with the performance of the SSA method: the quality of the RD solver has not been evaluated for this magnitude of $t_\text{max}$, and especially the poor results for $\gamma$ hint at a problem with the placement algorithm that might be fixed automatically once cell sizes are properly implemented, besides the before-mentioned coarseness in grouping. The excellent error measures for $D$ provide clues that in principle the SSA method should be capable of estimating the parameters even on the larger time scale.
Therefore we conclude that the \textit{in silico} validation of the SSA method regarding the biomarker $\sqrt{D/\gamma}$ proved successful, and under certain circumstances also regarding $\sqrt{D\cdot \gamma}$, delivering another puzzle piece towards their clinical application.

Apart from the mentioned possibilities for variation and improvement, there are yet other modifications we aim to investigate in the future: it is possible to extend the proposed PDE solver to three-dimensional spatial domains which allows for a three-dimensional discrete tumor model. A two-dimensional slice from such a digital tumor provides a more realistic base for the \textit{in silico} validation, as it does not necessarily cut through the center of the tumor which is a relatively optimistic assumption that was made in this work. Also it would be interesting to see how the SSA method performs with cutouts instead of whole slides.

\section*{Author Contributions}

\textbf{VH}: Conceptualization, numerical solution of the RD-equations and testing of the solver, development of the normalization method, \textit{in silico} biopsy generation, validation process. Writing: original draft, review \& editing.
\textbf{PS}: Consulting in and provision of the SSA method and related resources, consulting in the development of the normalization method and the validation process. Writing: review \& editing.
\textbf{JZ}: Numerical solution of the RD-equations, consulting in the \textit{in silico} biopsy generation. Writing: review \& editing.
\textbf{HE}: Conceptualization and co-supervision. 
\textbf{CK}: Project supervision. Writing: review \& editing.

\section*{Declarations}

\subsection*{Funding and Data Availability}
This work was supported by internal funds from the Technical University of Munich.

Pirmin Schlicke was in part funded within the APART-USA program of the OeAW.

Heiko Enderling and Pirmin Schlicke were supported in part by the JAYNE KOSKINAS TED GIOVANIS FOUNDATION FOR HEALTH AND POLICY, a Maryland private foundation dedicated to effecting change in the healthcare industry for the greater public good. The opinions, findings, and conclusions or recommendations expressed in this material are those of the author(s) and not necessarily those of the JAYNE KOSKINAS TED GIOVANIS FOUNDATION FOR HEALTH AND POLICY, its directors, officers, or staff.

The code used in this work is available at \url{https://github.com/veronikahofmann/2d_insilico_biopsy_ssa_validation}.

\subsection*{Declaration of generative AI and AI-assisted technologies in the writing process}

In this work, ChatGPT (\cite{chatgpt}) was used to support code writing and debugging. After using this tool, the authors reviewed and edited the content as needed and take full responsibility for the content of the publication. 

\subsection*{Competing Interests}

The authors have no competing interests to declare that are relevant to the content of this article.

\printbibliography

\appendix
\section{Derivation of the Two-dimensional Dirac-Delta Distribution in Polar Coordinates}\label{app:dirac_delta_derivation}

The i.c.~of \eqref{eq:2D_polar_RD_exp_growth} and \eqref{eq:2D_polar_RD_log_growth} is given by the two-dimensional Dirac-Delta function $\delta^2$ transformed into polar coordinates, which we will derive here from the one-dimensional Dirac-Delta distribution in cartesic coordinates. In one spatial dimension, the Dirac-Delta distribution is heuristically defined as
\begin{equation*}
    \delta (x) := \begin{cases}
        \infty & \text{if $x = 0$,}\\
        0 & \text{otherwise,}
    \end{cases}
\end{equation*}
such that $\int_{\mathbb{R}} \delta (x) \text{d}x = 1$.
From this integral condition, it follows for the two-dimensional Dirac-Delta distribution
\begin{align*}
    1 &= \int_{\mathbb{R}^2} \delta^2 (x_1, x_2) \:\text{d} x_1\text{d} x_2 \\
    &= \int_0^{2\pi} \int_0^\infty \delta^2(r) \: r \: \text{d}r \text{d} \phi = 2 \pi \int_0^\infty \delta^2(r)  \: r \:  \text{d}r,
\end{align*}
where the $r$ inside the integral is the volume element in polar coordinates, and the factor $2 \pi$ comes from the rotational symmetry of the $n$-dimensional Dirac-Delta distribution. 
The simplest choice for a function $\delta_a^2: \mathbb{R} \to \mathbb{R}$ approximating $\delta^2 \in \mathcal{C}( \mathbb{R}^2)$ in polar coordinates is therefore $\delta_a^2 (r) := \frac{1}{2\pi r}\delta_a(r)$, where $\delta_a: \mathbb{R} \to \mathbb{R}$ approximates the one-dimensional Dirac-Delta distribution.

\section{Numerical Modifications to Ensure Dirac-Delta Integral Condition}\label{app:dirac_delta_approx_mods}

In theory, all of the approximations in~\eqref{eq:dirac_delta_approximations} fulfill $\lim_{\varepsilon \to 0} f_j (r, \varepsilon) = \delta^2(r)$, and also the integral condition $\int_{\mathbb{R}^2} \delta^2 (x) \text{d}x = 1$, but of course this is not attainable numerically because the parameter $\varepsilon$ cannot be chosen sufficiently small. To help a bit, two measures are implemented:

\begin{itemize}
    \item In the case of $j=1, 2, 4$, 
    the first two spatial nodes $f_j(r_1)$ and $f_j(r_2)$ get assigned values $c_j$ that ensure the integral condition of the Dirac-Delta function. If $\varepsilon$ is chosen $>1$, this can lead to negative values of the initial condition at these first nodes. 
    To avoid this, we only allow parameter values $\varepsilon < 1$. Note that this was not necessary for $f_3$, since here we can govern the integral directly via the additional parameter $\varepsilon_*$.
    \item The support is made compact by setting the approximation to zero after some $r > r_\text{lim}$ to avoid perturbations. Without this measure the solutions $u$ computed with $f_2$ and $f_4$ have been observed to develop a second front of cells at the outer boundary $r_\text{max}$ of the domain, likely caused by the cell mass being slightly larger than 0 on the whole domain.
\end{itemize}
\begin{figure*}[htb]
\centering 
\includegraphics[width=\columnwidth]{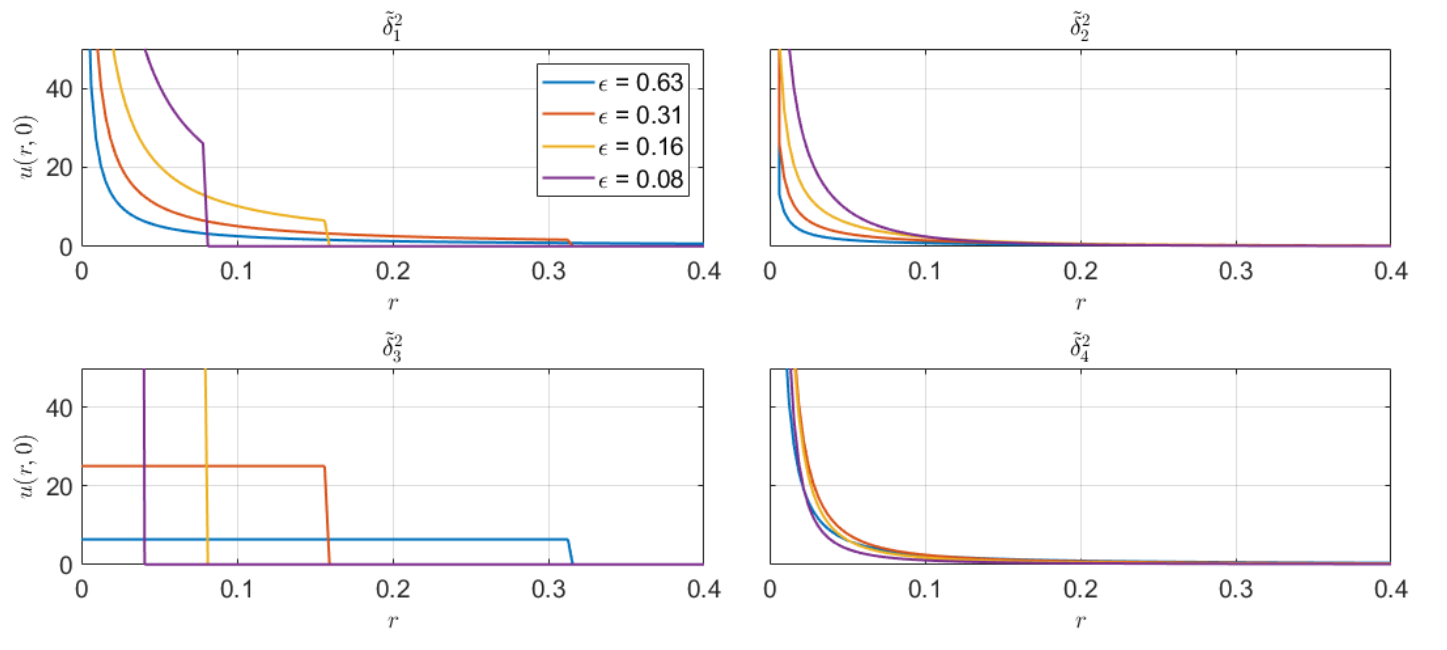}
\caption{\label{fig:dirac_delta_approx}Dirac-Delta approximations \eqref{eq:dirac_delta_approximations} for various parameter values $\varepsilon$.}
\end{figure*}
See Fig.~\ref{fig:dirac_delta_approx} for a depiction of the so-defined $\tilde\delta^2_j$. 
Note that there are many other possible Dirac-Delta approximations than the ones presented in~\eqref{eq:dirac_delta_approximations}. The current selection features the most prominent approximations which feature easy numerical integrability. Some approximations had to be excluded because they provoked singularities in the finite difference method, for instance periodic functions such as $f (r) = \frac{1}{2 \pi r} \cdot \frac{1}{2 \pi} \cdot \frac{\sin((1/\varepsilon + 0.5) r)}{\sin(0.5 r)}$.

\section{Rationale for the Normalization Method}\label{app:rationale_for_normalization}

For the simulation to produce meaningful results that are comparable for different parameter combinations, the objective is to ensure that normalized solutions of the RD-equation with exponential growth fulfill $u_\text{exp}/\| u \| \in [0, 1]$. 

Let $t > 0$ be fixed. Then, for the \say{norm} from \eqref{eq:unorthodox_norm} it holds 
\begin{equation*}
    \min_{r\in [0, r_\text{max}]} \frac{u_\text{exp}(r, t)}{\| u(r, t) \|} = \frac{\min_r u_\text{exp}(r, t)}{\| u(r, t) \|} = 0,
\end{equation*}
because by the b.c.~we have that $u(r_\text{max},t) = 0$, and it appears natural that $\| u(r, t) \| > 0$: for $t$ fixed, $\| \cdot \|$ considers the maxima of the solutions $u_\text{exp}$ and $u_\text{log}$. For both reaction terms, the RD-equation with $D>0$, $\gamma >0$ \say{flattens} and increases the initial cell mass, which is set as $1$ due to the i.c. The integral over the solution which defines this mass is $\geq 1$ for all $t\geq 0$, which is only possible if at least for some $r$, we always have $u>0$, and therefore $\max_r u>0$, regardless of the type of reaction term.

Let's consider the maximum, i.e.,
\begin{equation*}
    \max_{r\in [0, r_\text{max}]} \frac{u_\text{exp}(r, t)}{\| u(r, t) \|} = \frac{\max_r u_\text{exp}(r, t)}{\| u(r, t) \|}.
\end{equation*}
As $t>0$, we know $\max_r u_\text{exp}(r, t) < \infty$, hence let $m := \max_r u_\text{exp}(r, t)$. Then,
\begin{equation*}
    \| u(r, t) \| = \frac{m}{\max_r u_\text{log}(r, t)},
\end{equation*}
where we need to differ between the analytical and the numerical case:
\begin{itemize}
    \item Analytical: it is $\max_r u_\text{log}(r, t) \in (0,1]$ for $t > 0$. Then, 
    \begin{align*}
        \frac{\max_r u_\text{exp}(r, t)}{\| u(r, t) \|} &= \frac{m}{m/\max_r u_\text{log}(r, t)} = \max_r u_\text{log}(r, t) \in (0,1].
    \end{align*}
    \item Numerical: it is $\max_r u_\text{log}(r, t) \in (0,m]$ for $t > 0$. The solution of the logistic growth case can exceed 1 due to the aforementioned reason, and $m$ is the upper bound because of the regulatory effect of $1-u < 0$ acting on the exponential growth term $\gamma u$ as long as $u>1$. From here, the maximum of the norm depends on $t$:
    \begin{itemize}
        \item $t$ is small enough such that $\max_r u_\text{log}(r, t) \in (1, m]$: then, 
        \begin{equation*}
            \frac{\max_r u_\text{exp}(r, t)}{\| u(r, t) \|} = \max_r u_\text{log}(r, t) \in (1, m].
        \end{equation*}
        This is of course problematic as it does not ensure that the maximum value of the normalized solution is equal to 1. We thus need to choose the minimum $t$ that we use for the simulations large enough.
        \item $t$ is large enough so that $\max_r u_\text{log}(r, t) \in (0,1]$: then,
        \begin{equation*}
            \frac{\max_r u_\text{exp}(r, t)}{\| u(r, t) \|} = \max_r u_\text{log}(r, t) \in (0,1].
        \end{equation*}
    \end{itemize}
\end{itemize}
We conclude: in a scenario where the analytical solution to the RD-equation with logistic reaction term was available, the normalization method ensures cell densities between 0 and 1 for any point in time $t$. In the case at hand where we do not know the analytical solution, we need to choose $t$ large enough to have this condition fulfilled.

\section{Retrieval of $2\sqrt{D\cdot\gamma}$}\label{app:retrieval_of_Dtimesgamma}

In the case where we use no information about tumor age, we can only retrieve $\sqrt{D/\gamma}$. The reason for this is the non-unique solution of the RD-equation if $t$ is treated as a parameter: the solution \eqref{eq:analytical_sol_2D_exp} of \eqref{eq:2D_polar_RD_exp_growth} is the same for the parameter tuples $(D, \gamma, nt)$ and $(nD, n\gamma, t)$, $n\in \mathbb{R}$:
\begin{equation*}
    u_\text{exp} (r) = \frac{1}{4 n \pi D t} e^{-\frac{r^2}{4nDt} + n\gamma t}.
\end{equation*}
This behavior is shared with the numerical solution of \eqref{eq:2D_polar_RD_log_growth} (which can be seen in experiments, data not shown), hence the normalization with the maximum of $u_\text{log}$ does not provide additional information to help distinguish $(D, \gamma, nt)$ from $(nD, n\gamma, t)$. Since the factor $n$ cancels out in $\frac{nD}{n\gamma} = \frac{D}{\gamma}$, this is not a problem for the retrieval of $\sqrt{D/\gamma}$, but in the case of the second biomarker used in \cite{PirminsNewPaper}, $2\sqrt{D\cdot\gamma}$, we cannot make a unique estimate as it is $D\gamma \neq n^2D\gamma$.

\section{Plots for Tests A ii+B, C i, C ii}\label{app:solver_tests_additional_figs}

\begin{figure*}[htb!]
\centering 
\includegraphics[width=\columnwidth]{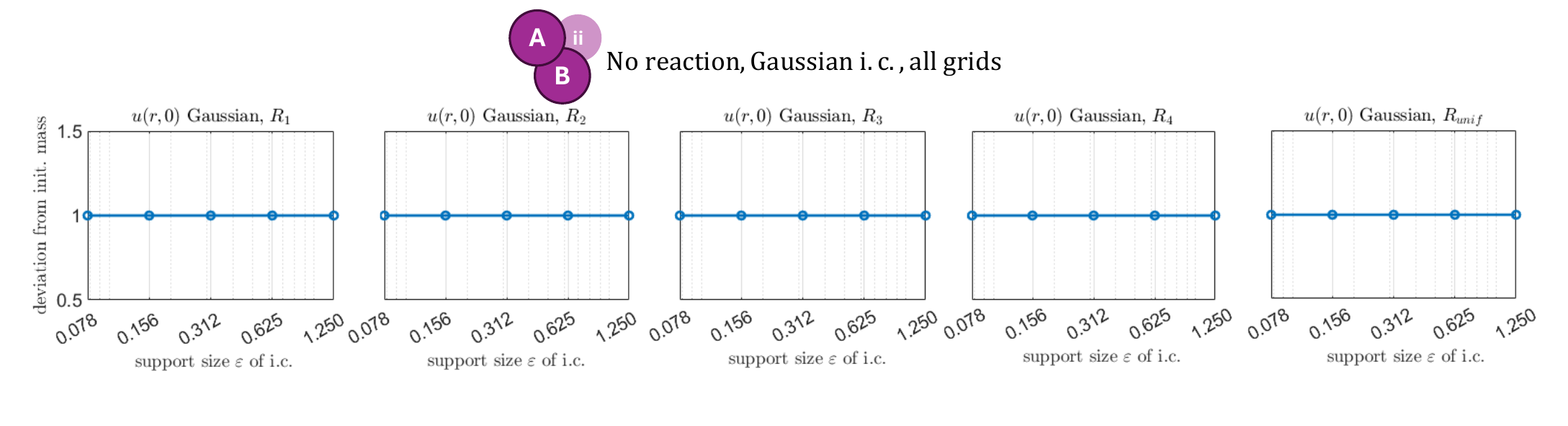} 
\caption{\label{fig:test_ABii}Mass conservation of the numerical solver with Gaussian i.c.~without reaction term on all grids. $D = 5\:000\frac{\mu\text{m}^2}{\text{day}}$ is selected because slow diffusion challenges the conservative properties of the solver the most, allowing to recognize deviations from the initial mass quickly. The initial mass is indicated as 1, deviations are measured proportionally.}
\end{figure*}

\begin{figure*}[htb!]
\centering 
\includegraphics[width=\columnwidth]{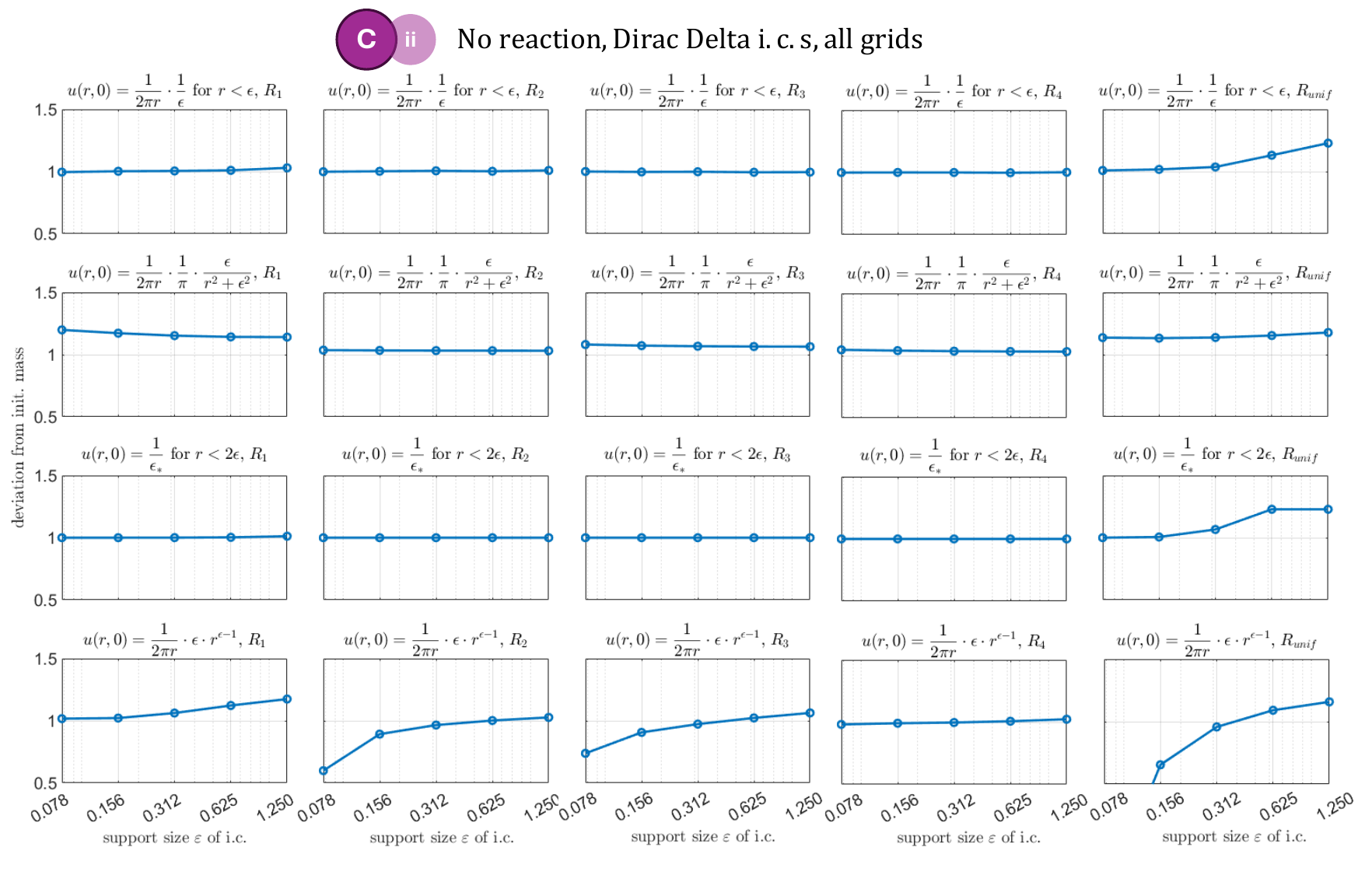}
\caption{\label{fig:test_Cii}Mass conservation of the numerical solver with the Dirac-Delta approximations from \eqref{eq:dirac_delta_approximations} as i.c.s~without reaction term on all grids. Again, $D = 5\:000\frac{\mu\text{m}^2}{\text{day}}$ is selected, and the initial mass is indicated as 1 with proportionally measured deviations.}
\end{figure*}

\begin{figure*}[htb!]
\centering 
\includegraphics[width=\columnwidth]{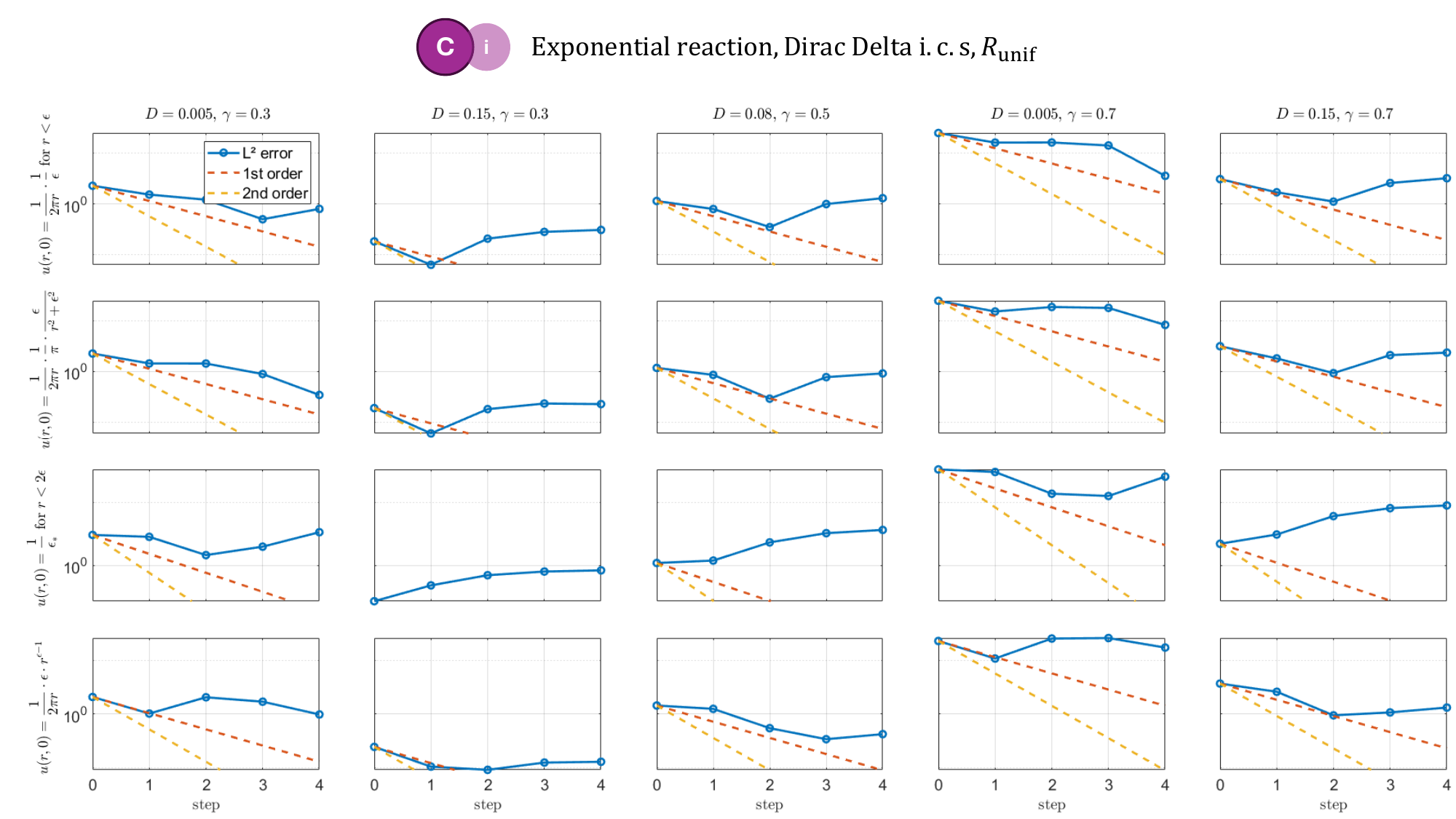} 
\caption{\label{fig:test_Ci_Runif}Convergence of the numerical solution with the Dirac-Delta approximations from \eqref{eq:dirac_delta_approximations} as i.c.s~with exponential reaction on $R_\text{unif}$. The i.c.s are preserved row-wise, the $(D, \gamma)$-combinations are preserved column-wise.}
\end{figure*}

\begin{figure*}[htb!]
\centering 
\includegraphics[width=\columnwidth]{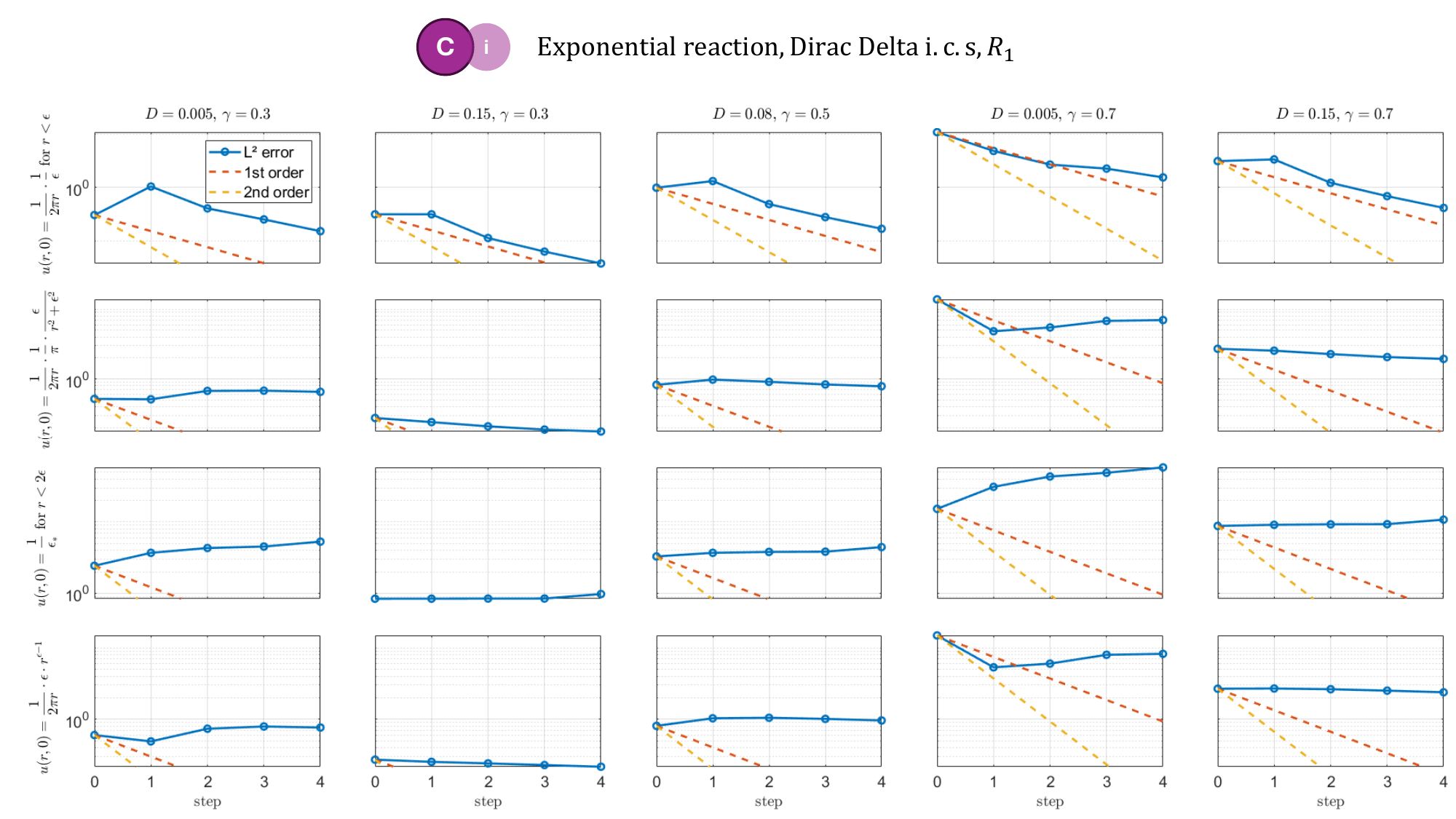} 
\caption{\label{fig:test_Ci_R1}Convergence of the numerical solution with the Dirac-Delta approximations from \eqref{eq:dirac_delta_approximations} as i.c.s~with exponential reaction on $R_1$.}
\end{figure*}

\begin{figure*}[htb!]
\centering 
\includegraphics[width=\columnwidth]{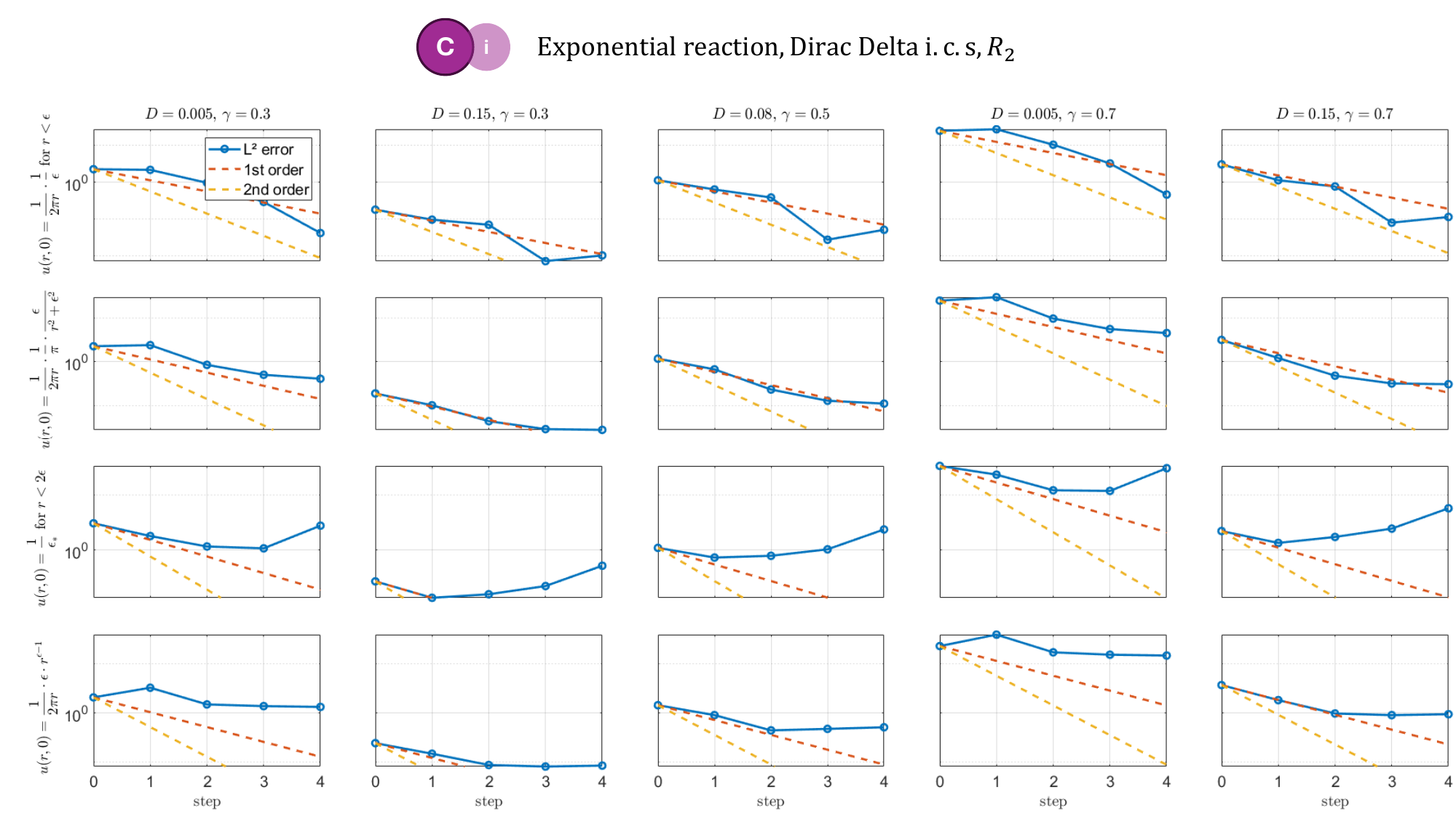} 
\caption{\label{fig:test_Ci_R2}Convergence of the numerical solution with the Dirac-Delta approximations from \eqref{eq:dirac_delta_approximations} as i.c.s~with exponential reaction on $R_2$.}
\end{figure*}

\begin{figure*}[htb!]
\centering 
\includegraphics[width=\columnwidth]{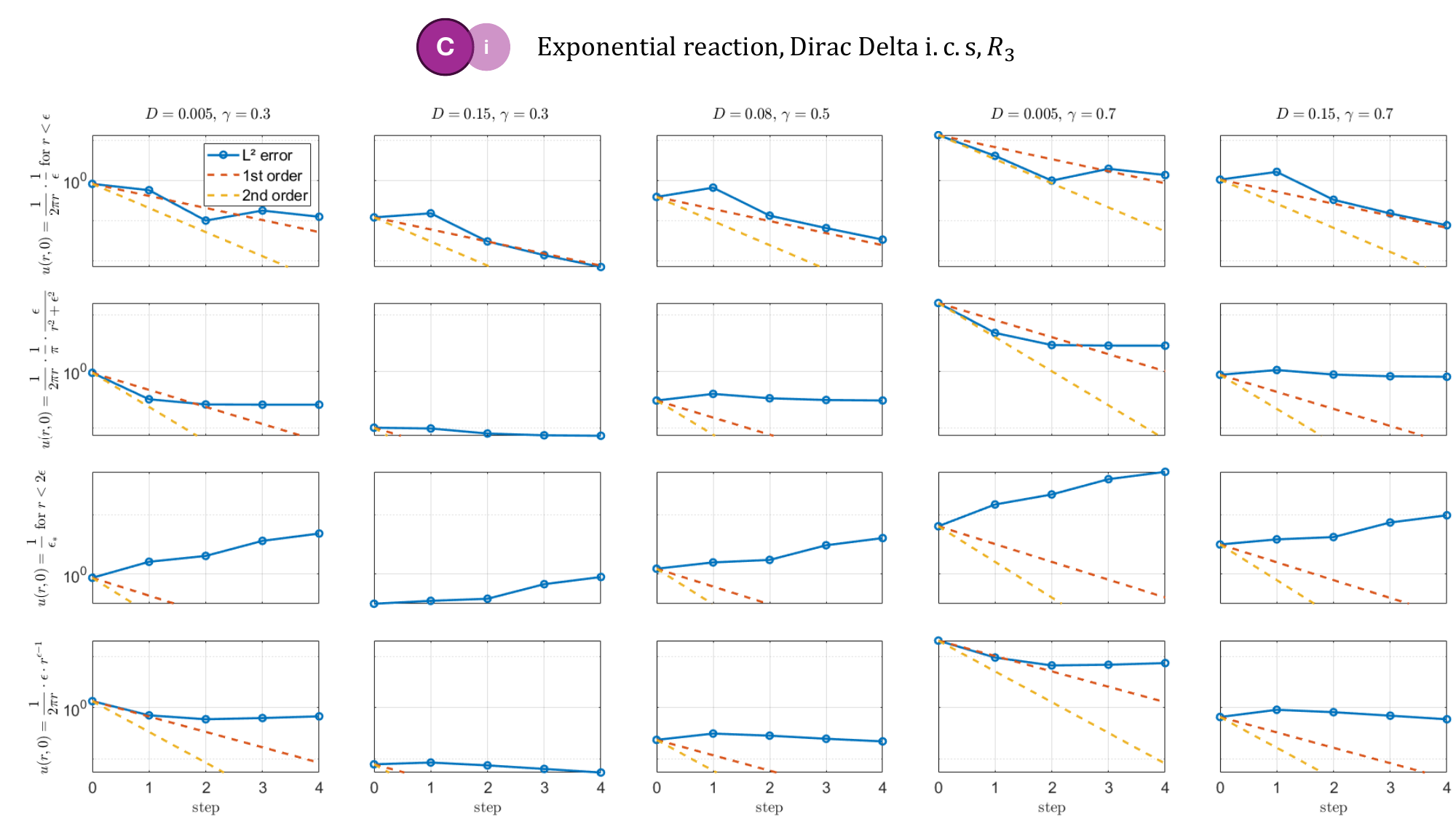}
\caption{\label{fig:test_Ci_R3}Convergence of the numerical solution with the Dirac-Delta approximations from \eqref{eq:dirac_delta_approximations} as i.c.s~with exponential reaction on $R_3$.}
\end{figure*}

\begin{figure*}[htb!]
\centering 
\includegraphics[width=\columnwidth]{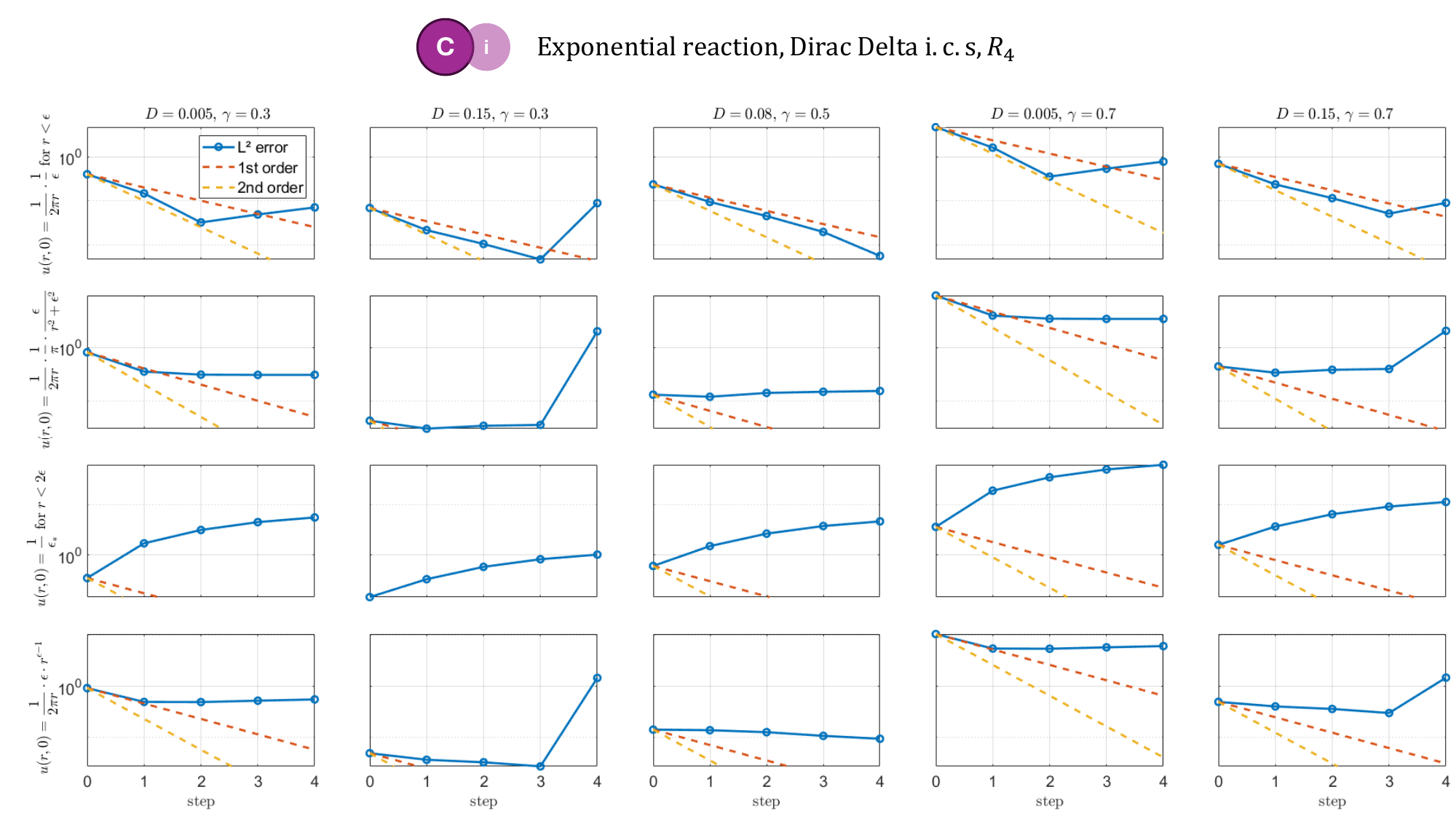} 
\caption{\label{fig:test_Ci_R4}Convergence of the numerical solution with the Dirac-Delta approximations from \eqref{eq:dirac_delta_approximations} as i.c.s~with exponential reaction on $R_4$.}
\end{figure*}

The results of the mass conservation test A ii, B are shown in Fig.~\ref{fig:test_ABii}. The results of the second mass conservation test C ii can be found in Fig.~\ref{fig:test_Cii}, and the results of the convergence test C i are depicted in Fig.s~\ref{fig:test_Ci_Runif}-\ref{fig:test_Ci_R4}.

\section{Cell Placement with Collision Check}\label{app:collision_cell_size}

\begin{figure*}[htb!]
\centering 
\includegraphics[width=\columnwidth]{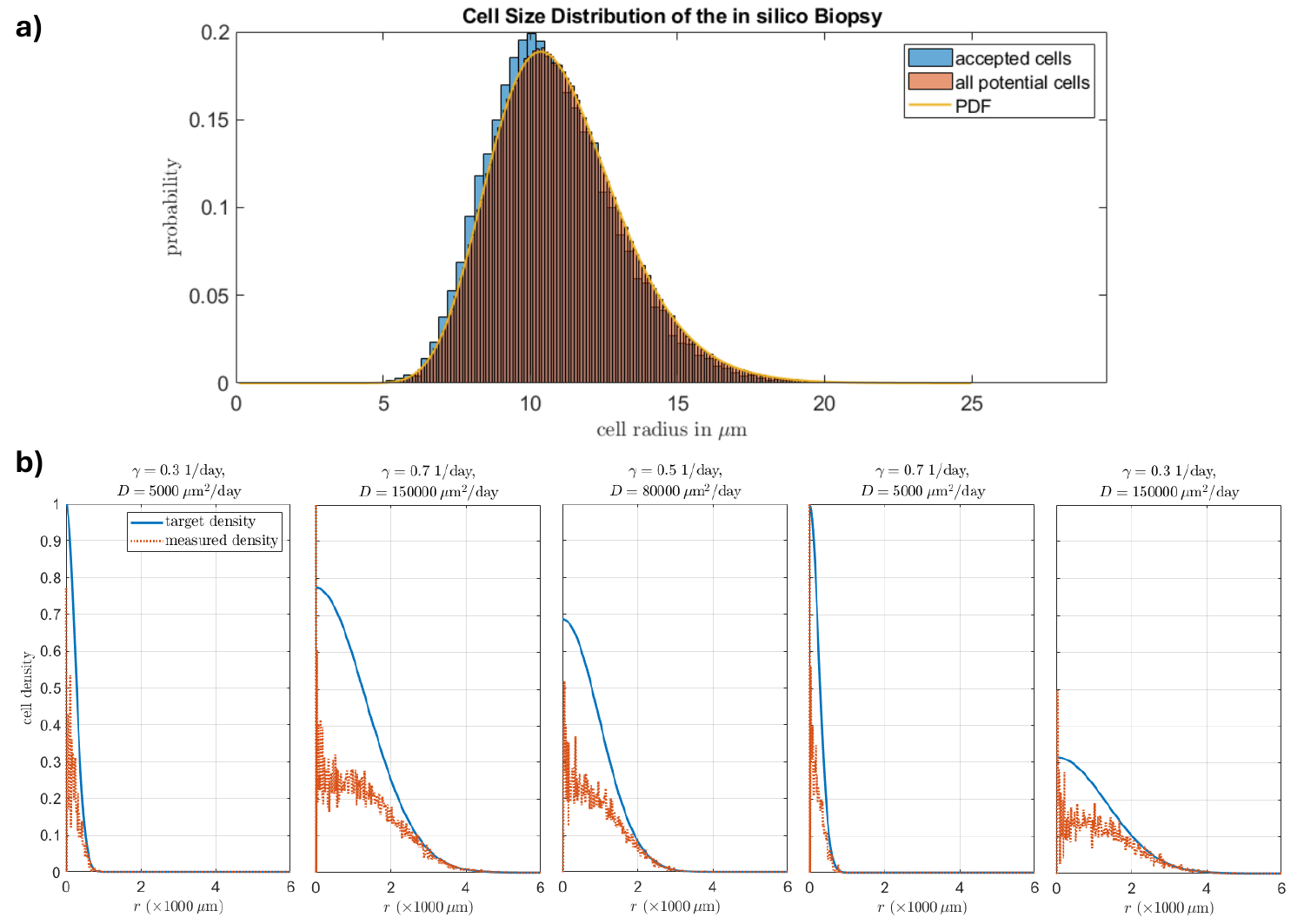} 
\caption{\label{fig:cellsize_invariance}Implications of the collision avoidance in the cell placement algorithm. \textbf{a)} Invariance of the cell size distribution under the selection algorithm. The depicted curve is the theoretical probability density function of the cell size distribution from which the simulation draws the samples. Parameters of the log-normal distribution: $\text{ln} (\theta) \sim \mathcal{N} (2.3779, 0.2)$ (corresponds to a mean cell radius $\theta$ of $11 \mu$m in linear scale). \textbf{b)} Desired density $\|u(r, t)\|$ versus measured cell density in the \textit{in silico} biopsy for different parameter combinations. While the fluctuations stem from the random generator and are expected, their mean deviation from $\|u\|$ is out of control, and the total cell mass in the biopsy is much lower than desired for all parameter choices. As in all tests regarding the numerical solver and the placement algorithm, $t_\text{max} = 6$ days is used.}
\end{figure*}

If one wants to implement incompressible cells with fixed sizes and collision avoidance in the placement of cells on the \textit{in silico} biopsy, one can perform a check as follows: before the new cells are definitely placed on the plane, collisions are avoided by checking the pairwise distances 1) within the cohort of new cells, and 2) with the cells that were placed in previous iterations. Also for cost efficiency regarding 2), distances are only checked for cells placed in some of the last iterations, their number determined by the value of $\sigma$. More precisely, the number of iterations that we need to trace back depends on $\sigma^2$: as for Gaussian distributed random variables it holds that $99.7\%$ of realizations can be found within a distance of three standard deviations $\sigma$ from the mean, we need to check the iterations $i-1, i-2, ...$ whose circles have radii $r$ with $r_i - r \leq 6 \sigma$. Why six instead of three? The \say{three standard deviation rule} also holds true for the cells that were placed in the previous iterations. Of course, the choice of three $\sigma$ is arbitrary, but a risk in this magnitude to place two cells too close to each other seems acceptable. The pairwise distances between the new points and the relevant previous ones are checked, and one partner per invalid pair is eliminated. Eventually, the new cells are filtered with the concentration $u(r_i)$: each of the remaining points has a likelihood of $u(r_i)$ to make it to the plane.

A main question in this context is how cell sizes will be implemented. In the biopsy, the cells are depicted as dots in a scatter plot, representing the locations of the cell nuclei, and their bodies are modeled by blocking enough space surrounding the dots. Every single solid tumor contains cancer cells of vast genomic heterogeneity, which is -- amongst other features -- expressed in size differences~(\cite{Mirzayans.2023}). Given that too similar cell sizes also might introduce artifacts to the statistical methods that will be applied to the \textit{in silico} biopsies, it is assumed that the best practice would be to introduce varying cell sizes, e.g.~following a log-normal distribution as suggested in the experimental setup section. 

The collision avoidance in the modified algorithm preserves the distribution of the cell sizes provided that a sufficient number of cells can be placed on the biopsy (their total number is mainly governed by the concentration $u$). As an example, consider Fig.~\ref{fig:cellsize_invariance} a): about a million cells were tested, and roughly $10\:000$ were eventually accepted because they kept enough distance to their neighbors and because they did not exceed the concentration that is required at their current position. The accepted cells follow the exact same distribution as the total number of tested cells because the selection by concentration follows a uniform likelihood, and the cell that is deleted because it is too close to its neighbor is selected uniformly as well. 

While these attempts to make the biopsy more realistic appear reasonable in theory, they turn out to lead to rather disappointing results, which is why they are not featured in the main text of this article. It appears that the selection algorithm that deletes colliding cells is so strict that it leads to a severe under-filling of space, as it can be seen in Fig.~\ref{fig:cellsize_invariance} b).
This is especially remarkable given that the formula by which the occupied area is calculated by design rather over- than underestimates the cell density.

\section{Tested Parameter Combinations}\label{app:D_gamma_combis}

Table~\ref{tab:D_gamma_combis} shows the $(D, \gamma)$-combinations that were tested for $t_\text{max} = 6$ days, and Tab.~\ref{tab:D_gamma_combis_larget} shows the combinations that were tested additionally to the combinations from Tab.~\ref{tab:D_gamma_combis} with $D\leq0.02$ for $t_\text{max} = 100$ days. Hence, in total 38 combinations were used for the short-term experiments, and 28 for the long-term experiments. They are chosen without any specific system, the only objective is to have them distributed over the $D$- and $\gamma$-domains.

\begin{table}[h!]
\centering
\begin{tabular}{|c | c || c | c || c | c || c | c || c | c || c | c |}
\hline
$D$ & $\gamma$ & $D$ & $\gamma$ & $D$ & $\gamma$ & $D$ & $\gamma$ & $D$ & $\gamma$ & $D$ & $\gamma$ \\ 
\hline\hline
0.005 & 0.3 & 0.02 & 0.4 & 0.05 & 0.3 & 0.08 & 0.3 & 0.12 & 0.3 & 0.15 & 0.3\\
0.005 & 0.4 & 0.02 & 0.4 & 0.05 & 0.45 & 0.08 & 0.45 & 0.12 & 0.4 & 0.15 & 0.4\\
0.005 & 0.45 & 0.02 & 0.45 & 0.05 & 0.5 & 0.08 & 0.5 & 0.12 & 0.45 & 0.15 & 0.55\\
0.005 & 0.5 & 0.02 & 0.55 & 0.05 & 0.55 & 0.08 & 0.55 & 0.12 & 0.5 & 0.15 & 0.6\\
0.005 & 0.55 & 0.02 & 0.6 & 0.05 & 0.6 & 0.08 & 0.6 & 0.12 & 0.55 &  & \\
0.005 & 0.65 & 0.02 & 0.65 & 0.05 & 0.65 & 0.08 & 0.65 & 0.12 & 0.65 & & \\
0.005 & 0.7 & 0.02 & 0.7 & 0.05 & 0.7 & 0.08 & 0.7 & & & & \\
\hline
\end{tabular}
\caption{$(D, \gamma)$-combinations used for the validation when $t_\text{max} = 6$ days is set in the simulation. Multiply $D$ with $10^6$ to obtain the values in $\mu \text{m}^2/\text{day}$.}
\label{tab:D_gamma_combis}
\end{table}

\begin{table}[h!]
\centering
\begin{tabular}{|c | c || c | c || c | c |}
\hline
$D$ & $\gamma$ & $D$ & $\gamma$ & $D$ & $\gamma$ \\ 
\hline\hline
0.008 & 0.3 & 0.012 & 0.3 & 0.017 & 0.3 \\
0.008 & 0.4 & 0.012 & 0.4 & 0.017 & 0.4 \\
0.008 & 0.5 & 0.012 & 0.5 & 0.017 & 0.5\\
0.008 & 0.6 & 0.012 & 0.6 & 0.017 & 0.6 \\
0.008 & 0.7 & 0.012 & 0.7 & 0.017 & 0.7 \\ 
\hline
\end{tabular}
\caption{$(D, \gamma)$-combinations additional to the ones with $D \leq 0.02$ in Tab.~\ref{tab:D_gamma_combis} used for the validation when $t_\text{max} = 100$ days is set in the simulation. Multiply $D$ with $10^6$ to obtain the values in $\mu \text{m}^2/\text{day}$.}
\label{tab:D_gamma_combis_larget}
\end{table}

\begin{figure*}[htb!]
\centering 
\includegraphics[width=\columnwidth]{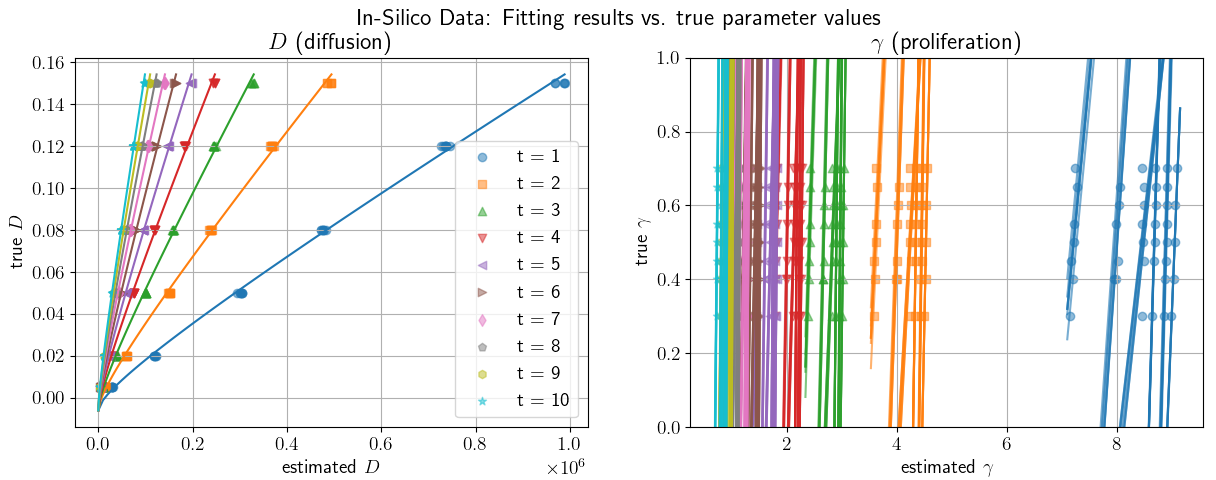} 
\caption{\label{fig:gamma_D_different_t}Plots of the conversion formulae for $\hat{D}$ and $\hat{\gamma}$ that were estimated using different $t_\text{max}$ in the PSD and 2PCF. Due to \eqref{eq:alt_conversion_formula_gamma}'s dependency on $f(\hat{D})$ which is unique for each $\hat{\gamma}$, each point $(\hat{\gamma}, \gamma)$ has its own fitted curve as this is a two-dimensional visualization of a three-dimensional hyperplane, even though the parameters $a, b, c$ are group-wise equal.}
\end{figure*}

\begin{figure*}[htb!]
\centering 
\includegraphics[width=\columnwidth]{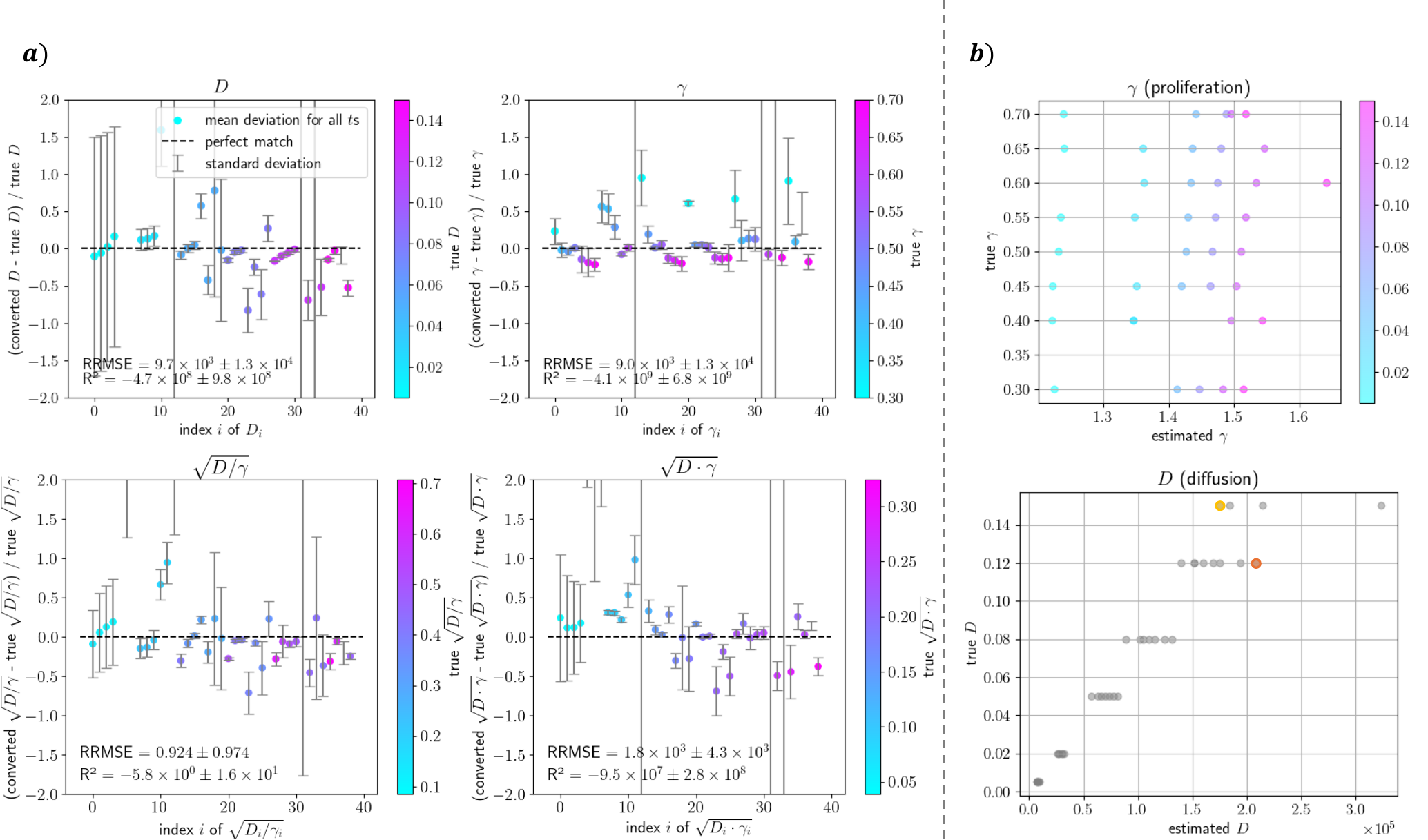} 
\caption{\label{fig:parity_log_growth}Relative residuals and comparisons for logistic growth simulations. \textbf{a)} Comparison of input parameter values $D, \gamma$ and converted output parameters $f(\hat{D})$, $g(\hat{\gamma}, \hat{D})$, using \eqref{eq:conversion_formula_D} and \eqref{eq:alt_conversion_formula_gamma}, and computing the output biomarkers as $\sqrt{f(\hat{D})/g(\hat{\gamma}, \hat{D})}$ and $\sqrt{f(\hat{D}) \cdot g(\hat{\gamma}, \hat{D})}$. The error bars illustrate the standard deviation between the relative residuals of all $t_\text{max}$. Note the scaling of the vertical axis: sometimes, the relative errors exceed $100\%$.
\textbf{b)} Input (true) parameters and output (estimate) parameters, estimated with exponential PSD and 2PCF formulae at a fixed point in time.}
\end{figure*}

\section{Details on the Validation}\label{app:validation_details}

To see how impactful it is to know which biopsies' $\hat{D}$ and $\hat{\gamma}$ were estimated using the same $t_\text{max}$ in the SSA method's curve fit, consider Fig.~\ref{fig:gamma_D_different_t}. The figure shows how the inverse quadratic relationship between $\hat{D}$ and $D$ becomes steeper as $t_\text{max}$ increases, and how $\hat{\gamma}$ is estimated larger for lower $t_\text{max}$-values (in the context of the short-term experiments). This $t_\text{max}$-dependent overestimation is likely a result from the too low $t_\text{max}$-values that are handed to the PSD and 2PCF: if a reaction-diffusion-PSD with $t_\text{max}=1$ is fitted to the results of a reaction-diffusion-simulation with $t_\text{max}=6$, the fit reacts to the advanced diffusion and proliferation of the cells by assuming larger $\hat{\gamma}$ and $\hat{D}$, compared to a fit where $t_\text{max}=4$ or $t_\text{max}=6$ is set.

These over- and underestimations are mitigated by having the information \say{same~$t_\text{max}$} as a reference point during the curve fit with \eqref{eq:conversion_formula_D} and \eqref{eq:alt_conversion_formula_gamma}, without needing to know the value of this shared $t_\text{max}$. The effectivity of this mitigation can be seen in Fig.~\ref{fig:parity_exp_growth} a) where the standard deviations of the converted parameters are negligible if they come from the same $(D, \gamma)$, but it can also be computed by evaluating the RRMSE and $\text{R}^2$ over all $t$-cohorts. Both quality of fit statistics remain very stable over time, as described in the Results section. 

\section{Validation with \textit{in silico} Biopsies Using Logistic Growth}\label{app:logistic_growth_validation}

When tumors grow for a while, their expansion will be limited by space and nutrients, leading to a switch from an exponential growth profile to a rather logistic one with carrying capacity. Simulating tumor growth with this kind of growth allows for a more realistic model of older tumors. As we have seen, it also facilitates the biopsy generation because it makes the normalization unnecessary.

Conducting the first kind of validation experiment with logistic instead of normalized exponential growth using the same $(D, \gamma)$-combinations as input as well as the exact same SSA procedure (again with high-confidence fitting of 2PCF and PSD, and $\text{R}^2>0.9$ for $90\%$ of the samples), we arrive at the results shown in Fig.~\ref{fig:parity_log_growth} a).

The errors are out of bounds, which must be due to the inappropriateness of conversion formulae. To see why this is the case, we need to go into more detail: different from earlier experiments with the normalized exponential growth biopsies, there is no time-independent consistency among conversions, visible in the large error bars. We have seen that for exponential growth, $\hat{D}$ is mapped to the same $f(\hat{D})$ independently of $t$, which is a consequence of a suitable conversion formula: consider the left-hand plot of Fig.~\ref{fig:gamma_D_different_t} and observe how close all points $(\hat{D}, D)$ are to their respective $t$-wise curves $D \approx f(\hat{D})$, not only vertically (i.e.,~the residuals), but also horizontally. The story looks entirely different for the logistic growth biopsies: the bottom plot of Fig.~\ref{fig:parity_log_growth} b) shows how this \say{horizontal clustering} is not the case here. This makes the definition of a well-defined conversion formula virtually impossible because there are estimates $\hat{D}_1$, $\hat{D}_2$ with $\hat{D}_1 < \hat{D}_2$ (imagine $\hat{D}_1$ to be the yellow dot, $\hat{D}_2$ the orange dot in Fig.~\ref{fig:parity_log_growth} b)). While $D$ and $\hat{D}$ are clearly positively correlated, i.e.,~if $D$ grows, then $\hat{D}$ is expected to grow as well, the marked points expose the exact opposite relationship. It is possible to mitigate these phenomena up to a certain point provided that the differences between $\hat{D}_1$ and $\hat{D}_2$ are not too large (as is sometimes necessary in the conversion of $\gamma$), but apparently in this experimental setup, this was not successful.

\end{document}